\documentclass[pre,twocolumn,showpacs,floatfix,superscriptaddress]{revtex4}

\usepackage{listings}
\usepackage{graphicx}
\usepackage{epstopdf}
\usepackage{multirow}
\usepackage{color}
\usepackage{amssymb,amsfonts,amsmath}
\usepackage[english]{babel}
\usepackage{float}

\newcommand{\mc}{\mathcal{C}}

\begin{document}

\title{Finding statistically significant communities in networks}

\author{Andrea Lancichinetti}\affiliation{Complex Networks \& Systems
  Lagrange Laboratory, ISI Foundation, Turin, Italy}
\affiliation{Department of Physics, Politecnico di Torino, Turin, Italy}

\author{Filippo Radicchi}\affiliation{Amaral Lab, Chemical and Biological Engineering,
Northwestern University, Evanston, IL, USA}

\author{Jos\'e J. Ramasco}\affiliation{Instituto de F\'{\i}sica
  Interdisciplinar y Sistemas Complejos IFISC (CSIC-UIB), E-07122
  Palma de Mallorca, Spain} 
\affiliation{Complex Networks \& Systems Lagrange Laboratory, ISI Foundation, Turin, Italy}

\author{Santo Fortunato}\affiliation{Complex Networks \& Systems Lagrange Laboratory, ISI Foundation, Turin, Italy}

\begin{abstract}
Community structure is one of the main structural features of 
networks, revealing both their internal organization and the similarity
of their elementary units. Despite the large variety of methods proposed
to detect communities in graphs, there is a big need for multi-purpose 
techniques, able to handle different types of datasets and the
subtleties of community structure. In this paper
we present OSLOM (Order Statistics Local Optimization Method), the first method capable to detect clusters in networks
accounting for edge directions, edge weights, overlapping communities,
hierarchies and community dynamics. It is based on the local
optimization of a fitness function expressing the statistical
significance of clusters with respect to random fluctuations, which is
estimated with tools of Extreme and Order Statistics. 
OSLOM can be
used alone or as a refinement procedure of partitions/covers
delivered by other techniques. We have also implemented sequential
algorithms combining OSLOM with other fast techniques, so that the
community structure of very large networks can be uncovered. 
Our method has a comparable performance as the best existing algorithms on artificial
benchmark graphs. Several applications on real networks are shown as well.
OSLOM is implemented in a freely available software ({\tt http://www.oslom.org}), and we
believe it will be a valuable tool in the analysis of networks.
\end{abstract}


\maketitle

\section{Introduction}

The analysis and modeling of networked datasets are probably the
hottest research topics within the modern science of complex 
systems~\cite{albert02,dorogovtsev01,newman03,pastor04,boccaletti06,barrat08,caldarelli07}.
The main reason is that, despite its simplicity, the network
representation can disclose some relevant features of the system at
large, involving its structure, its function, as well as the interplay
between structure and function. 
The elementary units of the system are reduced to simple points,
called {\it vertices} (or {\it nodes}), while their pairwise relationships/interactions are
pictured as {\it edges} (or {\it links}). It is fairly easy to spot
the two main ingredients of a graph in many instances. Therefore
networks can be found everywhere: in biology (e. g., proteins and their interactions),
ecology (e. g., species and their trophic interactions), society (e. g.,
people and their acquaintanceships). Other noteworthy examples include
the Internet (routers/autonomous
systems and their physical and/or wireless connections), the World Wide Web (URLs and their hyperlinks), etc..

The structure of most networks, beneath the intrinsic disorder due to
the stochastic character of their generation mechanisms, 
reveals a high degree of organization. In particular, vertices with similar 
properties or function have a higher chance to be linked to each
other than random pairs of vertices and tend to form highly cohesive
subgraphs, which are called {\it communities} (also {\it modules} or
{\it clusters}). Examples of communities are groups of mutual acquaintances in
social networks~\cite{girvan02,lusseau04,adamic05}, subsets of Web
pages on the same subject~\cite{flake02}, compartments in food webs~\cite{pimm79,krause03}, functional modules in
protein interaction networks~\cite{jonsson06}, biochemical pathways in metabolic
networks~\cite{holme03,guimera05}, etc.. 

Detecting communities in graphs may help to identify functional
subunits of the system and to uncover similarities among vertices that
are not apparent in the absence of detailed (non-topological)
information. Vertices belonging to the same community may be
classified according to their structural position within the cluster, which may be
correlated to their role. Vertices in the core of the cluster may have
a function of control and stability within the module, whereas boundary vertices 
are likely to be mediators between different parts of the graph.
The community structure of a network can also be a powerful visual representation
of the system: instead of visualizing all the
vertices and edges of the network (which is impossible on large
systems), one could display its communities and their mutual
connections, obtaining a far more compact and understandable
description of the graph as a whole.  
It is thus not surprising that community detection in graphs has 
been so extensively investigated over the last few years~\cite{fortunato10}.
A huge variety of different methods have been designed by a truly
interdisciplinary community of scholars, including physicists, computer
scientists, mathematicians, biologists, engineers and social scientists.

However, most algorithms currently available cannot handle 
important network features. Many methods are designed to find clusters
in undirected graphs, and cannot be easily (or not at all) extended to
directed graphs. However, there are many datasets for which edge
directedness is an essential feature. Citation networks, food webs and
the Web graph are but a few examples. Similar problems arise when edges
carry weights, indicating the strength of the interaction/affinity
between vertices, although extensions are generally easier in
this case. 

Likewise, the great majority of algorithms are not capable to deal
with the peculiar features of community structure. For example,
each vertex is typically assigned to a single cluster, while in several
instances, like in social networks, vertices are typically shared between two or more clusters.
In such cases communities are {\it overlapping} (and partitions become
{\it covers}) and very few methods account for this
possibility~\cite{baumes05,palla05,zhang07,gregory07,nepusz08,lancichinetti09,evans09,kovacs10},
which considerably increases the complexity of the problem.
Furthermore, community structure is very often {\it hierarchical},
i.e.  it consists of communities which include (or are included by)
other communities. Hierarchies are common in human societies and are crucial for an efficient management of large
organizations. Simon pointed out that hierarchy gives robustness and stability to complex
systems, yielding an evolutionary advantage on the long
run~\cite{simon62}. However, most community finding methods typically
look for the ``best'' partition of a network, disregarding the
possible existence of hierarchical structure. Instead, a method
should be able to recognize if there is hierarchical structure and, if yes,
identify the corresponding levels~\cite{sales07,clauset07,clauset08}.

It is also very important for a method to distinguish communities from
pseudo-communities. The existence of clusters indicate a
preference by some groups of vertices to link to each other. But, if
the linking probability is the same for all pairs of vertices, like in
random graphs, no communities are expected. In this case, concentrations of edges
within groups of vertices are simply the result of random fluctuations, they do not
represent potentially non-trivial structures. Many algorithms are not
able to see this difference and find clusters in random graphs as
well, although they are not meaningful. Scholars have just begun to
assess the issue of significance of
clusters~\cite{bianconi09,lancichinetti10}.

Finally, given the recent availability of time-stamped networked
datasets, it is now possible to carry out quantitative studies on the
dynamics of community structure,
about which very little is known~\cite{hopcroft04,backstrom06,chakrabarti06, palla07,asur07,mucha10}. 
A simple way to treat dynamic datasets is to analyze snapshots of the
system at different times separately, and then map communities of
different snapshots onto each other, such that one can follow the
dynamic of each cluster in time. However, focusing on individual
snapshots means disregarding the
information on the system at previous times. Ideally a partition/cover
of the system at time $t$ should be faithful both to its structure at time
$t$ and to its history~\cite{chakrabarti06,mucha10}. 

In this paper we propose the first method able to meet all
requirements listed above, the Order Statistics Local Optimization
Method (OSLOM). It is a method that optimizes locally the statistical
significance of clusters, defined with respect to a global null model. The
concept of statistical significance is inspired by recent work of some
of the authors~\cite{lancichinetti10,radicchi10}. 
The paper is structured as follows. After introducing the method, we test its
performance on artificial benchmark graphs, comparing it with
the performances of the best algorithms currently available. Next, we pass to the analysis of 
real networks, followed by a final discussion on the work. Some of the
tests on artificial and real networks are reported in the Appendix.

\section{Methods}

\subsection{Statistical significance of clusters}

In this section we explain how to estimate the statistical
significance of a given cluster. OSLOM will use the significance as a
fitness measure in order to evaluate the clusters. 
Following our previous work~\cite{lancichinetti10}, we define it as 
the probability of finding the cluster in a random null model, i. e.
in a class of graphs without community structure. 
We choose the configuration model~\cite{molloy95} as our null model. 
This is a model designed to build random networks with a given 
distribution of the number of neighbors of a vertex (degree).
The networks are generated by joining randomly vertices under the
constraint that each vertex has a fixed number of neighbors, taken
from the pre-assigned degree distribution. This is basically the same null model adopted by Newman
and Girvan to define modularity~\cite{newman04b}. 

\begin{figure} [h!]
\begin{center}
\includegraphics[width=\columnwidth]{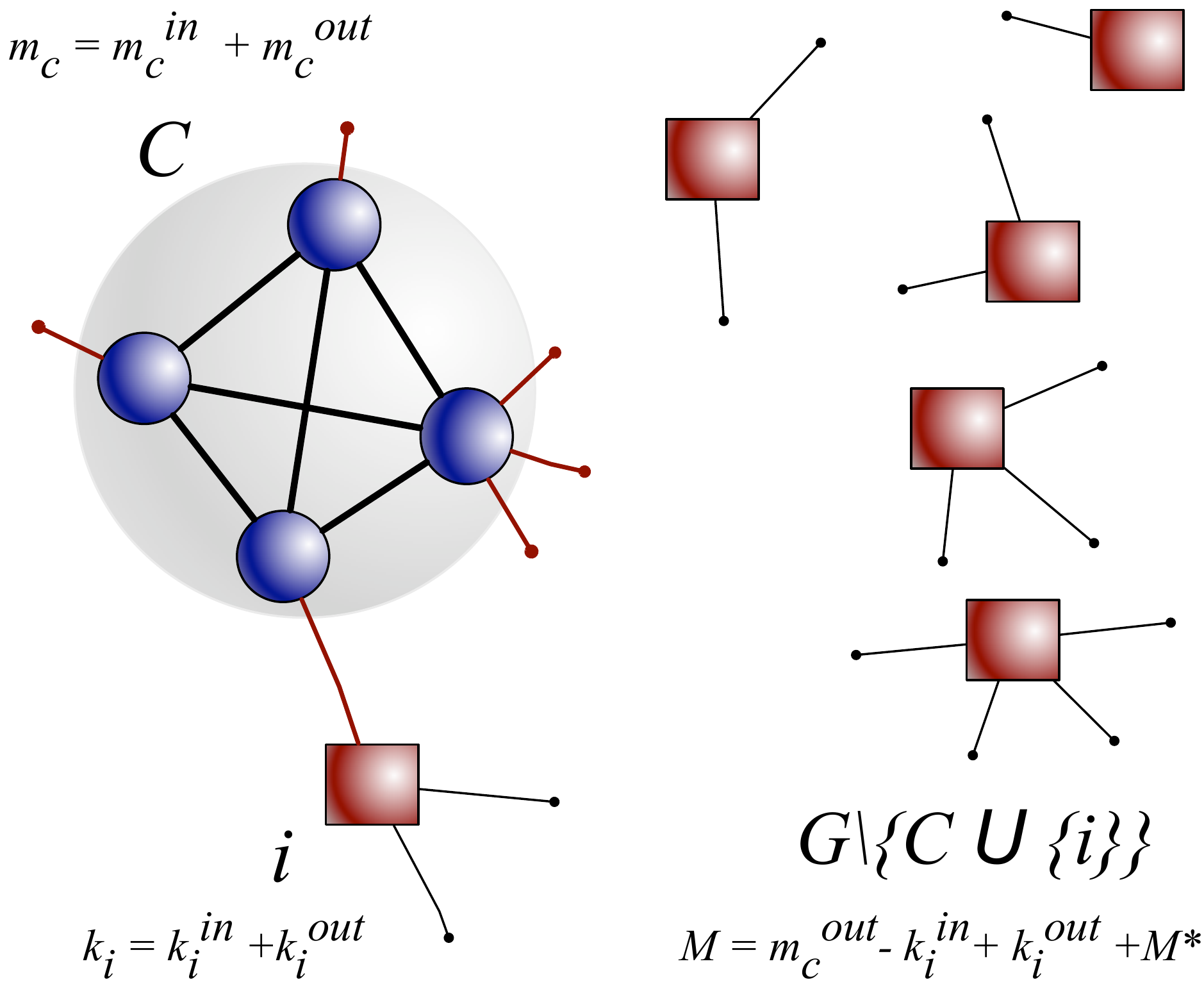}
\caption{A schematic representation of a subgraph $\mc$,
whose significance is to be assessed. 
The subgraph $\mc$ is embedded 
within a random graph generated by the configuration
model. The degrees of all vertices of the network are fixed, in
the figure we have highlighted the degrees of $\mc$
($m_\mc$), of the vertex $i$ at the center of the analysis ($k_i$) and of the
rest of the graph ${\mathcal G} \setminus [\,\mc \cup
\{i\}\,]$ ($M$). These quantities are expressed as sums of contributions
which are internal to their own set of vertices (as $M^*$) or related
to subgraph $\mc$ (in or out). This notation is used in the
distribution of Eq. 1.}
\end{center}
\end{figure}

We start from a graph ${\mathcal G}$ with $N$ vertices and $E$ edges.
The framework for the analysis is sketched in
Fig. 1. We are given a subgraph $\mc$, whose significance
is to be assessed, a vertex $i \notin \mc$
and the degree of the vertices of the rest of the graph ${\mathcal G} \setminus [\,\mc \cup
\{i\}\,]$.
The degree of subgraph $\mc$ is $m_\mc$, $k_i$ is the degree of $i$,
and the rest of vertices have a total degree $M$. We can separate the
above quantities in the contributions internal or external to $\mc$
($m_\mc^{in}, m_\mc^{out}, k_i^{in}$ and $k_i^{out}$); the internal
degree of ${\mathcal G} \setminus [\,\mc \cup
\{i\}\,]$ is $M^*$ (Fig. 1). 

Let us suppose that $\mc$ is a subgraph of graphs 
generated by the configuration model, where each vertex maintains
the degree it has on the graph ${\mathcal G}$ at study. 
We assume that the internal degree $m_\mc^{in}$ of the subgraph is fixed. 
If all the other edges
of the network are randomly drawn, the probability that $i$ has
$k_i^{in}$ neighbors in $\mc$ can be written as~\cite{radicchi10} 
\begin{equation}
\label{hyper_eq1}
p(k_i^{in}|i,\mc,{\mathcal G})= A \frac{2^{-k_i^{in}}}{k_i^{out}! \; k_i^{in}!\; (m_{\mc}^{out}- k_i^{in})! \; (M^*/2)!}.
\end{equation}
This equation enumerates the possible configurations of the network
with $k_i^{in}$ connections between $i$ and $\mc$. 
The factorials of the formula express the multiplicity of
configurations with fixed values of $k_i^{in}$, $k_i^{out}$, $(m_{\mc}^{out}- k_i^{in})$
and $M^*/2$, whereas the power of $2$ in the numerator stays for the
multiplicity coming from the permutation of the extremes of edges
lying between $i$ and $\mc$.
Several of the terms in the expression can actually be written as
a function of constants 
and $k_i^{in}$, such as $k_i^{out} = k_i-k_i^{in}$ and $M^* = 2\,E  -
m_\mc -m_\mc^{out} - 2\, k_i +2k_i^{in}$.  
The normalization factor $A$ includes terms not depending on $k_i^{in}$ and ensures that
\begin{equation}
\sum_{k_i^{in}: {M^*}\geq 0} p (k_i^{in}|i,\mc,{\mathcal G}) = 1.
\end{equation}
Further details on the numerical implementation of the formula in
Eq.~\ref{hyper_eq1}, 
as well as on the different approximations taken and their limits, are
included in Appendix~\ref{a1}.

The probability of Eq.~\ref{hyper_eq1} provides a tool to rank the
vertices external to $\mc$ according 
to the likelihood of their topological relation with the group. If vertex $i$ shares many more edges with the vertices of subgraph
$\mc$ than expected in the null model, we could consider the inclusion
of $i$ in $\mc$, 
since the relationship between $i$ and $\mc$ is ``unexpectedly" strong. 
In order to perform the ranking 
the cumulative probability $r(k_i^{in}) =
\sum_{j=k_i^{in}}^{k_i}p(j|i, \mc, {\mathcal G})$ of having a number 
of internal connections equal or larger than $k_i^{in}$ is estimated, following Ref.~\cite{lancichinetti10}. Given that the
vertex degree is a discrete variable, the cumulative distribution has
a specific step-wise
profile for each value of $k_i$. In order to facilitate the comparison
of vertices with different degrees, 
we implement a bootstrap strategy by assigning to each vertex $i$ a
value of $r$, $r_i$, 
randomly drawn from the interval $[r(k_i^{in}),
r(k_i^{in}+1)]$. This choice is important for a
meaningful estimate of the clusters' significance; other options
(e. g., taking the middle points of the interval) could lead to
the identification of meaningful clusters in random graphs. The
bootstrap introduces a stochastic 
element in the assessment procedure, which will, in turn, lead to the use of Monte Carlo techniques.

The variable $r$ bears the information regarding the likelihood of
the topological relation of each vertex with $\mc$ and has
an important feature: it is a uniform random variable distributed between zero and
one for vertices of our null model graphs. Calculating its 
order statistic distributions is thus a relatively easy task. The first candidate 
among the external vertices to be part of $\mc$ is the vertex with the
lowest value of $r$, that we indicate $r_1$. The cumulative distribution of $r_1$
in the null model is then given by
\begin{equation}
\label{cnstar}
\Omega_1(r) = P(r_1<r) = 1 - (1- r)^{N-n_\mc},
\end{equation}
where $n_\mc$ is the number of vertices in $\mc$. In general, let $r_{q}$ be the value of variable $r$ with rank 
$q$ (in increasing order of the variable $r$).

\begin{figure} [h!]
\begin{center}
\includegraphics[width=\columnwidth]{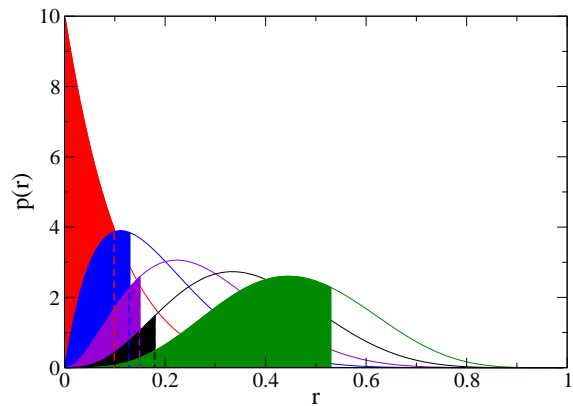}
\caption{Probability distributions of the scores $r$ of vertices
external to a given subgraph $\mc$ of the graph. The score $r_{q}$ is
the $q$-th smallest score of the external vertices. In this particular case
there are $10$ external vertices.  In the figure, we plot $p(r_{1})$,
$p(r_{2})$,  $p(r_{3})$, $p(r_{4})$,  $p(r_{5})$ (from left to right). As an example, the shaded areas show the cumulative probability
$\Omega_q$ for a few values of $r$ that would correspond to the values
estimated in a practical situation. 
In this case, the black area, $q= 4$, is the least extensive and so
$c_m = \Omega_4$. 
If $\phi(c_m)<P$, the vertices with scores $r_1$,$r_2$, $r_3$ and $r_4$ will be added to $\mc$.}
\end{center}
\end{figure}
Its cumulative
distribution is (Fig. 2):
\begin{equation}
\label{equation_c}
\Omega_q(r)= p(r_ {q} <x) = \sum _{i= q }^{N-n_\mc}
{N-n_\mc\choose{i}} x^i (1-x)^{N-n_\mc-i} .
\end{equation}

The reason for the use of order statistics is that we
assume that clustering methods tend to include in each community those
vertices which are most strongly connected to vertices of the community. 
Due to correlations (the vertices in the clusters tend to be connected), we cannot 
calculate the statistics of the internal connections to the clusters,
but we can do it safely 
for the external vertices. The values of the different $\Omega_q$ inform 
us of how much the external vertices of a group are compatible with
the statistics expected in the null model. 
To evaluate the full group, we define $c_m= \min_q \{\Omega_q(r_q)\}$ among all the 
neighbors of $\mc$, where $r_q$ are their corresponding ranked values for the $r$ variable. 
The distribution of $c_m$ can be easily tabulated numerically since it
only depends on $N-n_\mc$.  The cumulative distribution will be
denoted as $P(c_m < x) = \phi(x, N-n_\mc)$. 
In the following, we call $\phi(c_m,N-n_\mc)$ the {\it score} of the
cluster $\mc$. 

\subsection{Single cluster analysis}

Now that a score to evaluate the statistical significance of the clusters has been introduced, the next
step is to optimize the score across the network 
by dividing it into proper clusters. We describe first the optimization of
a single cluster score and will extend 
later the method to deal with the full network. First of all one has
to give the method a certain 
tolerance, in the following referred to as $P$. This parameter 
establishes when a given value of the score is considered significant. 
Our procedure consists of two phases: first, we explore the
possibility of adding 
external vertices to the subgraph $\mc$; second, non-significant vertices in $\mc$
are pruned.  They are described below and illustrated schematically in Fig. 3.

\begin{figure} [h!]
\begin{center}
\includegraphics[width=\columnwidth]{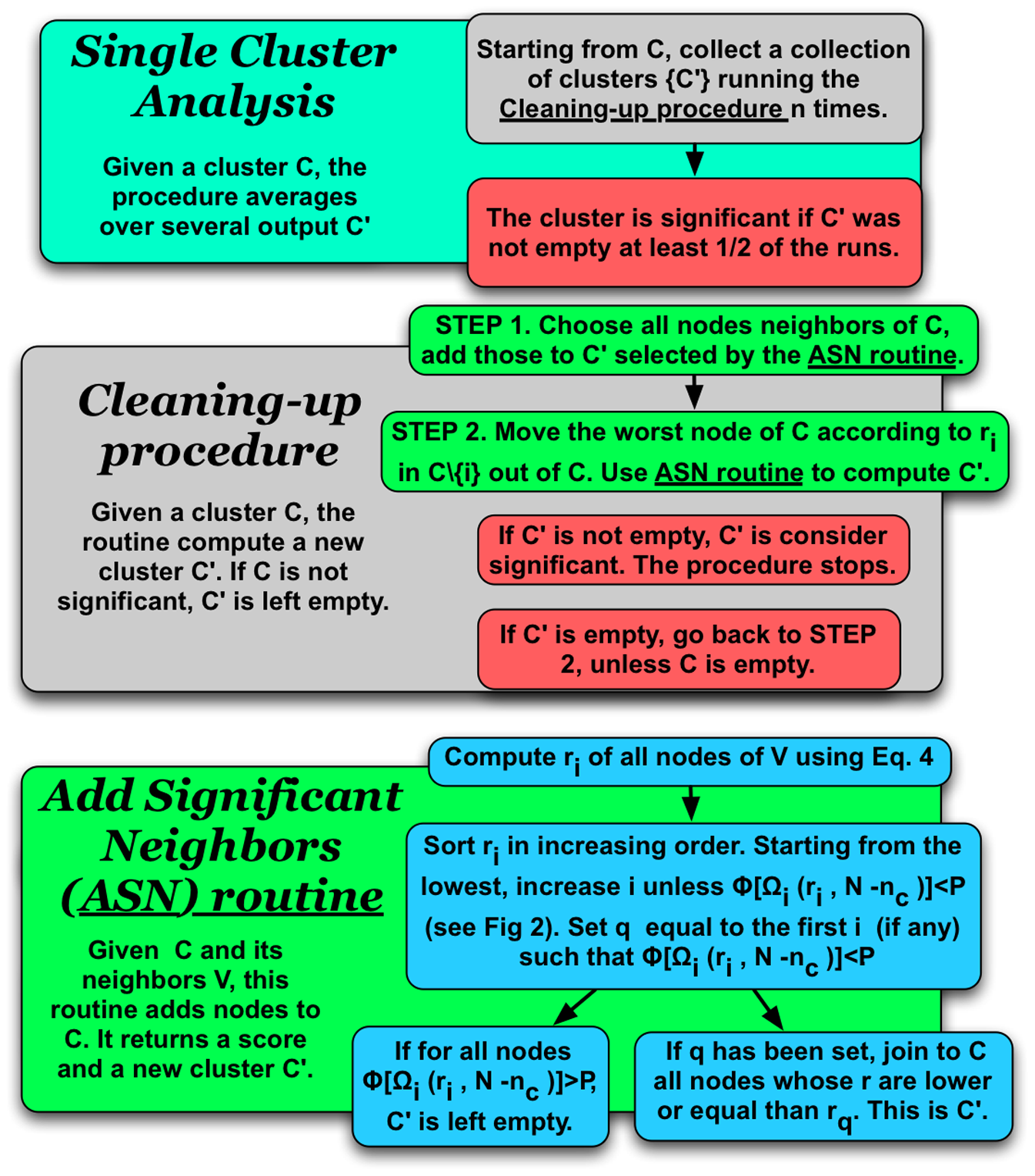}
\caption{Schematic diagram of the single cluster analysis.}
\end{center}
\end{figure}

\begin{enumerate}
\item{For each vertex $i$ outside $\mc$ and connected to
it by at least one edge the variable $r$ is computed. Then we calculate
$\Omega_1(r)$ for the vertex with the smallest $r$, by using Eq.~\ref{cnstar}. 
If $\phi(\Omega_1(r),N-n_\mc)<P$, we add the corresponding vertex to the
subgraph, which we now call $\mc^\prime$. If $\phi(\Omega_1(r),N-n_\mc)>P$,
one checks the second
best vertex, the third best vertex, etc.. If there is finally a vertex, say the $q$-th best
vertex, for which $\phi(\Omega_q(r),N-n_\mc)<P$, one includes all $q$
best vertices into subgraph $\mc$, yielding subgraph $\mc^\prime$. 
At this point, no other vertex outside $\mc$ deserves to enter the community 
since all the external vertices are compatible with the statistics of
the random configuration model. It may also happen that the inequality
$\phi<P$ above holds for no external vertex, in which case we add no
vertices to $\mc$ and $\mc^\prime=\mc$. Either way, we pass
to the second stage with the subgraph $\mc^\prime$.} 
\item{For each
vertex $i$ in $\mc^\prime$ the variable $r_i$ with 
respect to the set $\mc^\prime \setminus \{i\}$ is estimated. We pick the ``worst'' vertex $w$ of the cluster, i. e. the vertex
with the highest value of $r_i$. To check for its significance we
repeat step 1 for the subgraph $\mc^\prime \setminus \{i\}$. If $w$
turns out to be significant, we keep it inside $\mc^\prime$ and the analysis of the
cluster is completed. Otherwise, $w$ is moved out of $\mc^\prime$ and
one searches for the worst internal vertex of $\mc^\prime \setminus
\{w\}$. At some point we end up with a cluster $\mc^*$, whose internal
vertices are all significant and the process stops.} 
\end{enumerate}

The two-steps procedure is a way to ``clean up'' $\mc$. A cluster is left
unchanged only if all the external vertices 
are compatible with the null model and all the internal vertices are not. A few remarks are important here:
\begin{itemize}
\item There can be both \textit{good} vertices outside $\mc$ and
  \textit{bad} ones inside. It is important to perform the complete
  procedure described above, which
guarantees that the final cluster is significant with respect to the
present null model (see also Ref.~\cite{lancichinetti10}). 
\item 
The procedure is not deterministic, because of the
stochastic component in the computation of the cumulative probability $r$.
So one shall repeat all the steps several times. The cluster analysis
may deliver a subgraph $\mc^\prime$, 
in general different from $\mc$, or an empty subgraph.
For each vertex $i$ we compute the participation frequency $f_i$,
defined as the ratio between the number of times $i$ belongs to any
non-empty $\mc^\prime$ and the total number of iterations leading to
non-empty subgraphs. In general, we consider the
subgraph $\mc$ to be a 
significant cluster if the single cluster analysis yields a non-empty subgraph $\mc^\prime$ in more 
than $50\%$ iterations.
The final ``cleaned'' cluster includes those vertices for which
$f_i>0.5$.

\item In the worst-case scenario, the complexity of the cluster analysis scales with the number of vertices 
of $\mc$, times the number of neighbors of $\mc$, times the number of
loops needed to have reliable values for the $f_i$'s. The situation
can be considerably improved by keeping
track of the order of the external vertices at each step (using suitable data
structures) and by computing the score only for some reasonably good
vertices. For instance, one could pick just those vertices
with $r<0.1$. We numerically checked that changing
this threshold does not affect the results, but leads to a faster algorithm.

\end{itemize}

\subsection{Network analysis}

The previous procedure deals with a single cluster $\mc$.
It finds the external significant vertices 
and includes them into $\mc$. It also prunes those internal vertices that
are not statistically relevant. 
Now we extend this procedure by introducing an algorithm
able to analyze the full network. 
In order to do so, we follow the method proposed by some
of the authors in 
Ref.~\cite{lancichinetti09}. The starting point is a single vertex,
taken at random, 
in the absence of any information. Let us suppose that we start from a
random vertex $i$ and that our first group is $\mc = \{i\}$. The
method proceeds as follows:
\begin{enumerate}
\item $q$ vertices are added  
to $\mc$, considering the most significant among the neighbors of
the cluster. The number $q$ is taken from a distribution, which in principle
can be arbitrary. We choose a power law with exponent $-3$.
\item Perform the single cluster analysis. 
\end{enumerate}
We repeat the whole procedure starting from several vertices in
order to explore different regions of the network. 
This yields a final set of clusters that may
overlap. Such type of local optimization was originally implemented in the
Local Fitness Method~\cite{lancichinetti09}, to handle
overlapping communities. The algorithm stops when it keeps 
finding {\it similar} modules over and over. Ideally one wishes to
encounter the exact same clusters repeatedly. However, 
the stochastic element introduced when calculating the vertex score
can lead vertices, whose score is 
close to the threshold, to change their group assignments from one
realization to another. 
This can be a problem when we are trying to decide whether two groups
in different 
instances correspond to the same cluster. As a practical rule, we say that two groups $\mc_1$ and $\mc_2$ are
similar if $|\mc_1 \cup \mc_2|  /    \min(|\mc_1|, |\mc_2|)  > 0.5$,
in which case they deserve further attention. 
Indeed, it turns out that many of the clusters found are very similar
or combinations of each other. 
This leads to a very important question: given a set of significant
clusters, which ones should be kept? 

Let us consider the problem of choosing between two clusters $\mc_1$
and $\mc_2$ and the union of the two, $\mc_3$. 
A solution is to consider the subgraph ${\mathcal G}_3$ of the
vertices in $\mc_3$ and see 
if $\mc_1$ and $\mc_2$ are significant as modules of ${\mathcal G}_3$.
Strictly speaking we 
consider $\mc_1'$ and $\mc_2'$ which are the cleaned up clusters within
$G_3$ (i.e. with respect to subgraph ${\mathcal G}_3$ only,
neglecting the rest of the network). We 
discard $\mc_3$ if $|\mc_1' \cup \mc_2'| > P_2 \cdot |\mc_3|$, where
we set $P_2=0.7$. 
Otherwise we discard  $\mc_1$ and $\mc_2$ and we keep the union
$\mc_3$. Instead, if we 
have to decide among a set of $k$ clusters and their union, the
condition to prefer the submodules is $\cup_i \mc_i' > P_2 \cdot |\mc_u|$.

In general, we check if each cluster has significant submodules,
by looking for modules 
in the subgraph given by the cluster and using the condition above to
decide which ones to take. 
This leads to a set of significant minimal clusters, where
minimal means that they 
have no significant internal cluster structure, according to the condition
above. We also need 
to check whether unions of such minimal clusters do have internal
cluster structure, according to our rule, to decide whether the
clusters have to be kept separated or merged.
After doing this, we still end up with many \textit{similar}
modules. Given a pair of similar modules (in the sense defined above), we first check if 
their union has significant cluster structure: if it does
not, we merge the two clusters, otherwise we systematically prefer the
bigger one (if they are equal-sized, we pick the cluster with smaller
score).

After the completion of this procedure, the output is a cover of the
network. To reduce the stochasticity 
introduced by the bootstrap, the procedure is repeated in order to obtain
several covers. All clusters of the covers 
are analyzed as described 
above to select among them the ones which will appear in the final output.

\subsection{OSLOM}

We have described the cleaning of a single cluster and how the full
network is analyzed. In the following, all the ingredients 
are assembled together to form the algorithm that we call OSLOM (Order
Statistics Local 
Optimization Method).

\begin{figure} [h!]
\begin{center}
\includegraphics[width=\columnwidth]{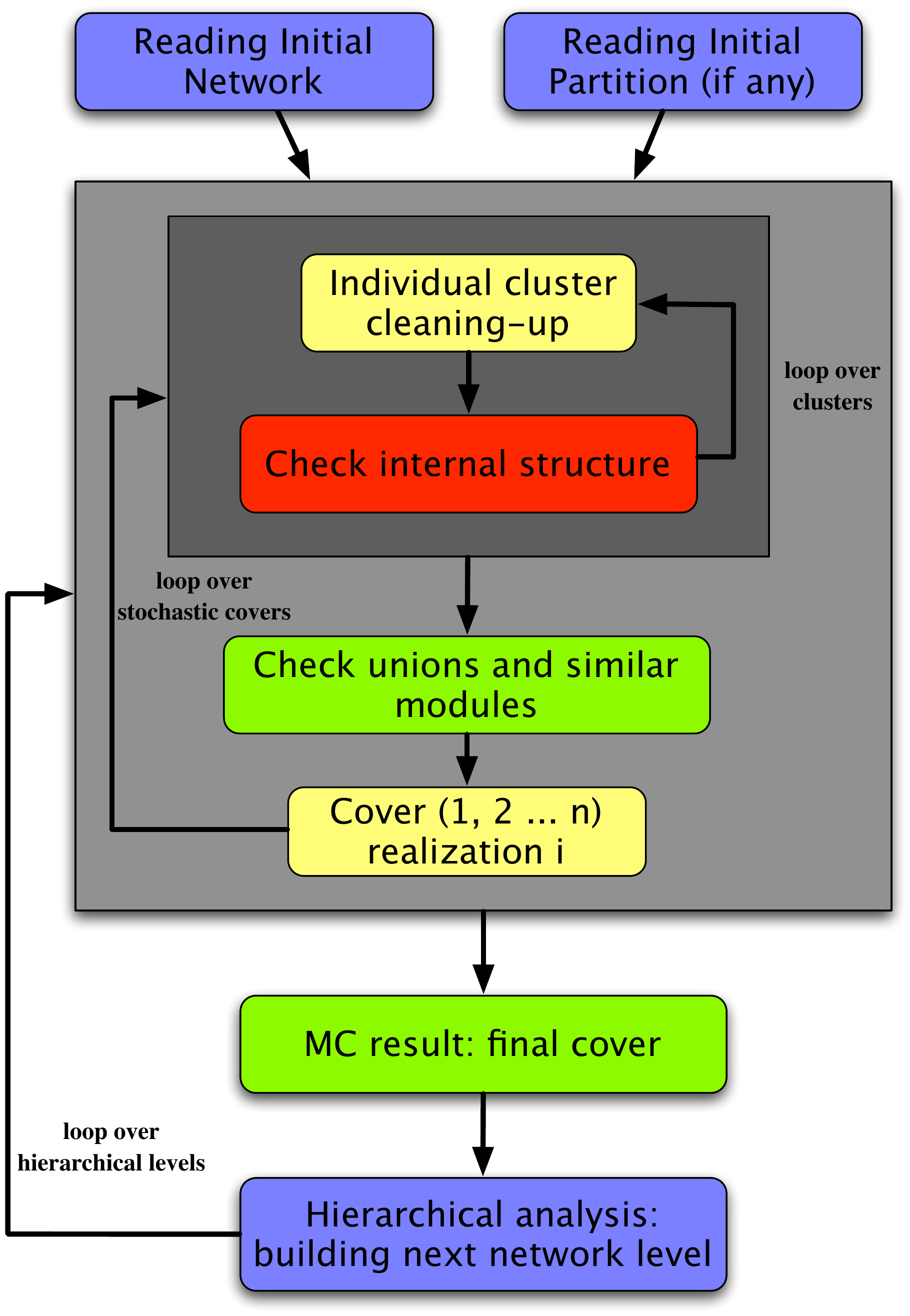}
\caption{Flux diagram of OSLOM. The levels of grey of the squares
represent different loop levels. One can provide an initial
partition/cover as input, from which the algorithm starts operating,
or no input, in which case the algorithm will build the clusters about
individual vertices, chosen at random.
OSLOM performs first a cleaning procedure of the
clusters, followed by a check of their internal structure and by a decision on
possible cluster unions. This is repeated with different choices of
random numbers in order to obtain better statistics and a more
reliable information. 
The final step is to generate a super-network for the next level of the hierarchical analysis.  }
\end{center}
\end{figure}
A flux diagram summarizing how it works can be
seen in Fig. 4. OSLOM consists of three phases:
\begin{itemize}
\item{First, it looks for significant clusters, until convergence;} 
\item{Second, it analyzes the resulting set of clusters, trying to
    detect their internal structure or possible unions thereof;}
\item{Third, it detects the hierarchical structure of the clusters.}
\end{itemize}
To speed up the method, one can start from a given partition/cover delivered by
another (fast) algorithm or from 
\textit{a priori} information. In those cases, the first step will be to clean up the given clusters. 

Once the set of minimal significant clusters has been found, the analysis of the hierarchies consists of the following steps. We
construct a new network formed by clusters, where each cluster is turned
into a supervertex and there are edges between supervertices if 
the representative clusters are linked to each other. The resulting superedges are 
weighted by the number of edges between the initial clusters. 
There is the problem of properly assigning edges between clusters, if
the edges are incident on overlapping vertices. 
Suppose to have an edge whose endvertices $i$ and $j$ belong to 
$\nu_i$ and $\nu_j$ clusters, respectively. This edge lies
simultaneously between any pair of clusters $\mc_i$ and $\mc_j$, with
$\mc_i$ including $i$ and $\mc_j$ including $j$. The contribution of
the edge to the superedge between $\mc_i$ and
$\mc_j$ equals $1/(\nu_i \cdot \nu_j)$. The resulting non-integer
weights may lead to non-integer values for the weight of superedges,
whereas we need integer values in order to use
Eq.~\ref{hyper_eq1}. For this reason, the weight of each superedge is
rounded to the nearest integer value. We stress that the weight we
deal with here indicates just how to ``split'' edges, it is not
related to the weight that edges may carry. If the original network is
weighted, the rescaled weight of an edge is $w/(\nu_i \cdot \nu_j)$,
$w$ being the weight of the edge in the network.
Once the supernetwork has been built,
one applies the method again, obtaining the second hierarchical
level. The latter is turned again into a supernetwork, as we explained
above, and so on, until the method produces no clusters. 
In this way OSLOM recovers
the hierarchical community structure of the original graph. 

We will describe next the main features of OSLOM, and what it adds
to the state of the art in community detection.

\subsubsection{Significant clusters}

The main characteristic of OSLOM is that it is based on a fitness
measure, the score, that is tightly related 
to the significance of the clusters in the configuration model. In
fact, the single cluster analysis 
is designed to optimize the cluster significance as defined in
Ref.~\cite{lancichinetti10}. 
Therefore the output of OSLOM consists of clusters that are unlikely
to be found in an equivalent random graph with the same
degree sequence. The tolerance $P$, fixed initially, determines whether such clusters are ``unexpectedly
unlikely'', and therefore significant, or not. 
So, if the method is 
fed with a random graph, the output will include very few clusters or even none at all.

\subsubsection{Homeless vertices}

The vertices in a random network will be deemed as homeless. Homeless
vertices are those that are not assigned to any cluster. 
This is a very important feature that OSLOM includes. The presence of random
noise or non-significant vertices is an issue that may occur in many real systems. 
However, very few clustering techniques take into account this possibility. 
In OSLOM, it comes as a natural output. We will quantitatively analyze
this feature when we test the method on benchmark graphs.

\subsubsection{Overlapping communities}

A natural output of OSLOM is the possibility for clusters to overlap.
Since each cluster is ``cleaned'' independently of the others, 
a fraction of its vertices may belong also to other clusters, eventually. We will
show the efficiency of OSLOM in unveiling overlapping vertices in suitably designed benchmarks.
 
\subsubsection{Cluster hierarchy}

Another relevant feature of OSLOM is the analysis of the
hierarchical structure of the clusters. 
As mentioned above, the third phase of our method includes a procedure
to take care of this issue. 
The results are very good on hierarchical benchmarks.

OSLOM generally finds different depths in different
hierarchical branches. In fact, 
when the algorithm is applied not all
vertices are grouped, as some of them are homeless. The coexistence of
homeless vertices with proper clusters yields a hierarchical structure
with branches of different depths.

\subsubsection{Weighted networks}
\label{wnetworks}

OSLOM can be generalized to weighted
graphs as well. We assume that the contributions to the probability of having a
connection between two vertices $i$ and $j$ with a certain weight $w_{ij}$, given the
vertex degrees $k_i$ and $k_j$ and their strengths, $s_i$ and $s_j$, is separable in two different
terms in the configuration model: one for the topology and another for the
weight~\cite{radicchi10}. 
The strength of a vertex is defined as the sum of the weights of all
the edges incident on it. We approximate the weight contribution by 
\begin{equation}
\label{p_w_ij}
p(w_{ij} > x |k_i,k_j,s_i,s_j)= \exp( - x / \langle w_{ij} \rangle),
\end{equation}
where $\langle w_{ij} \rangle = 2  \langle w_{i} \rangle  \langle
w_{j} \rangle /  (\langle w_{i} \rangle +  \langle w_{j} \rangle)$ is
the harmonic mean of the average weights of vertices $i$ and $j$,
defined as $ \langle w_{i} \rangle = s_i/k_i$ and $ \langle w_{j}
\rangle = s_j/k_j$, respectively. The idea behind this expression is that the weight of an edge of the null model should be 
proportional to the average weight of its endvertices.
We proposed the harmonic average because it is more sensitive to the small values of $\langle w \rangle$.

We use this distribution to define a new variable $r_w$, accounting for the probability of having a 
certain weight on a given edge with the strengths of the vertices and the general weight 
distribution known. We combine this variable $r_w$ with its
topological counterpart, $r_t$, obtaining a new variable $r_{wt}$. 
This is a non-trivial task since both probabilities are defined on a
different set of elements (see the Supporting Information S1). For $r_{wt}$
we can estimate, as before, the order statistic distributions
and we proceed just as we do for unweighted graphs.

\subsubsection{Directed graphs}
\label{directed}

OSLOM can be easily generalized to handle directed graphs. For that, we need to
define two uniformly distributed random variables $r_{out}$ and
$r_{in}$. The former is based on the probability that vertex $i$ has
outgoing edges ending on vertices of the given subgraph $\mc$,
the latter is based on the probability that $i$ has incoming edges
originating from vertices of $\mc$.
These two probabilities are computed through analogous formulas as in
Eq.~\ref{hyper_eq1} 
or numerical approximations to it.
The final score of vertex $i$ is given by the product ${r}_{in} \cdot
r_{out}$. We are able to calculate the distribution of this product
and therefore to estimate its order statistics (just as for the
weighted case, see Section 1.1. of Supporting Information S1).
The rest of the clustering method proceeds as explained above.
If graphs have edges with both directions and weights, we have four variables
for each vertex: $r_{in}$, $r_{out}$ and the corresponding versions
for the weights. The final score is given
again by the product of these four variables.

\subsubsection{Dynamical networks}

Time-stamped networked datasets are usually divided into snapshots,
condensing the relational information between vertices within different
time windows. Snapshots are typically analyzed separately, whereas it
would be more informative to combine the information from different
time slices. For instance, consider two snapshots
${\mathcal G}_t$ and ${\mathcal G}_{t+\Delta t}$ at times $t$ and
$t+\Delta t$, respectively. A simple idea is to find the partition/cover of the
network at time $t$, by applying the method to the corresponding
snapshot, and to use the result as an input for the application of the
method to the network at time $t + \Delta t$.
In this way one can see how the community structure at time
$t$ ``evolves'' to that at time $t+\Delta t$. This is a rather
general approach, it can be adopted for other algorithms for community
detection, like greedy optimization techniques. 
OSLOM has the useful property that it can start from any initial
partition/cover, which can be given as input. In this way
the clusters found in ${\mathcal G}_t$ can be used as
initial condition for the analysis of ${\mathcal G}_{t+\Delta t}$. With this approach, the new
partition/cover is closer 
to that in ${\mathcal G}_t$ and we are able to track the groups' evolution. 
Naturally, if the two snapshots are very different from each
other (because they refer to times between which the system has
changed considerably, for instance), OSLOM produces a partition/cover 
in ${\mathcal G}_{t+\Delta t}$ that is uncorrelated with that of
${\mathcal G}_t$.

\subsubsection{Complexity}
\label{complex}

The complexity of OSLOM cannot be estimated exactly, as it depends on the
specific features of the community structure at study. Therefore we carried out a
numerical study of the complexity, whose results are shown in
Fig. 5.

\begin{figure} [h!]
\begin{center}
\includegraphics[width=\columnwidth]{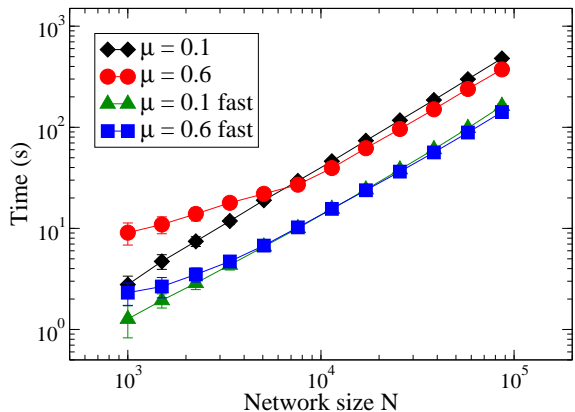}
\caption{Complexity of OSLOM. The diagram shows how the execution time
of two different implementations of the algorithm scales with the
network size (expressed by the number of vertices), for LFR benchmark graphs.}
\end{center}
\end{figure}

We apply the method on the LFR benchmark~\cite{lancichinetti08}, that we
have used extensively to test the performance of OSLOM. We have used both the standard version of the algorithm
and a fast implementation, in which the algorithm acts on the
partition delivered by a quick method. For each
version we have considered undirected and unweighted LFR benchmark
graphs with two different levels of mixtures between the clusters
($\mu=0.1$ and $\mu=0.6$, corresponding to well separated and well mixed clusters).
The other parameters needed to build
the LFR benchmark graphs are the same as
for the graphs used in Fig. 6.
The diagram of Fig. 5 shows the execution time (in seconds) as a function of the
number $N$ of vertices of the graphs. The processes were run on 
a workstation HP Z800.
The time scales as a power law
of $N$ with good approximation, if the graphs are not too small. The
behavior seems to depend neither on how mixed communities are,
nor on the particular implementation of the algorithm (there seems to
be just a factor between the corresponding curves). Power law fits of
the large-N portion of the curves yield an exponent $1.1(1)$, which
implies that the complexity is essentially linear in this case. 

\section{Results}


\subsection{Artificial networks}
\label{bench}

In this section we test OSLOM against artificial
benchmarks, comparing its performance with those 
of the best algorithms currently available. We mostly adopted the LFR
benchmark~\cite{lancichinetti08,lancichinetti09b},
a class of graphs with planted community structure and heterogeneous
distributions of vertex degree and community size. Tests on the well known
Girvan-Newman (GN) benchmark~\cite{girvan02} are shown in the
Supporting Information S1.
In this section we present tests on undirected and unweighted
networks, with and without hierarchical structure and overlapping communities. We also show how
OSLOM handles the presence
of randomness in the graph structure.
Tests on weighted networks and on directed networks can be
found in the Supporting Information S1.

In the following sections, for each network, we compose the
results of 10 iterations for the network analysis for
the first hierarchical level and the results of 50 iterations for higher levels, 
if any. The single cluster analysis was repeated 100 times for each cluster.

\subsubsection{LFR benchmark}
\label{lfr}

The LFR benchmark~\cite{lancichinetti08,lancichinetti09b}, like the GN
benchmark, is a particular case of the {\it planted $\ell$-partition
model}~\cite{condon01}, which is the simplest possible model of
networks with communities. The planted $\ell$-partition
model is a class of graphs whose vertices 
are divided into $\ell$ equal-sized groups, such
that the probability that two vertices of the same group are linked is
$p$, while the probability that two vertices of different groups are
linked is $q$, with $p>q$. 
The planted $\ell$-partition model is too simple to describe real
networks. Vertices have essentially the same 
degree and communities have the same size, at odds with empirical
analysis showing that both features typically are broadly
distributed~\cite{albert00,newman04,
  palla05,radicchi04,clauset04,lancichinetti10c}.
Therefore we have recently proposed a generalization of the model, the
LFR benchmark, by
introducing power-law distributions for the vertex degree and the community
size, with exponents
$\tau_1$ and $\tau_2$, respectively~\cite{lancichinetti08}. 
The LFR benchmark poses a far harder
challenge to algorithms than the benchmark by Girvan and Newman, which is regularly
used in the literature, and
is more suitable to spot their limits. We are of course aware that the
communities of the model are still too simple to match the communities
of real networks. Other features should be introduced, to
tailor the model graphs onto the real graphs. This is certainly
doable, and could be specialized to the particular domain of
applicability one is interested in. Still, the clusters of
the LFR benchmark are a much better proxy of real communities than the
clusters of other benchmark graphs.

Vertices of the LFR benchmark have a fixed degree
(in this case taken from the given power law distribution), so the 
two parameters $p$ and
$q$ of the planted $\ell$-partition model are not independent and 
we choose as independent variable the {\it mixing parameter} $\mu$, 
which is the ratio of the number of external neighbors of a vertex by
the total degree of the vertex. Small values of $\mu$ indicate well separated
clusters, whereas for higher and higher values communities become more
and more mixed to each other.

\begin{figure}
\begin{center}
\includegraphics[width=\columnwidth]{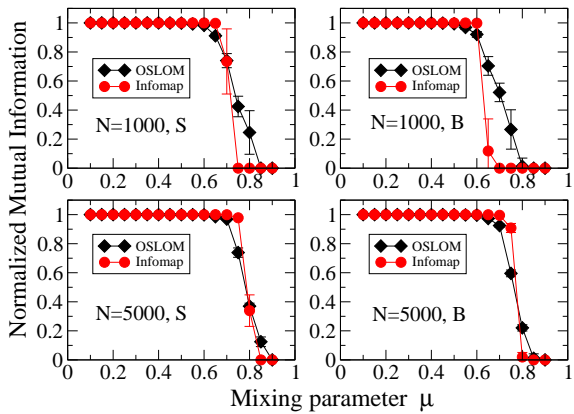}
\caption{Tests on undirected and unweighted LFR benchmark graphs
  without overlapping communities. The parameters of the
  graphs are: average degree $\langle k\rangle=20$, maximum degree
  $k_{max}=50$, exponents of the power law distributions are
  $\tau_1=2$ for degree and $\tau_2=1$ for community size, S and B mean that community sizes are in the range
  $[10, 50]$ (``small'') and $[20, 100]$ (``big''), respectively. We
  considered two network sizes: $N=1000$ (top) and $N=5000$ (bottom). 
  The two curves refer to OSLOM (diamonds) and Infomap (circles).}
\end{center}
\end{figure}
\begin{figure}
\begin{center}
\includegraphics[width=\columnwidth]{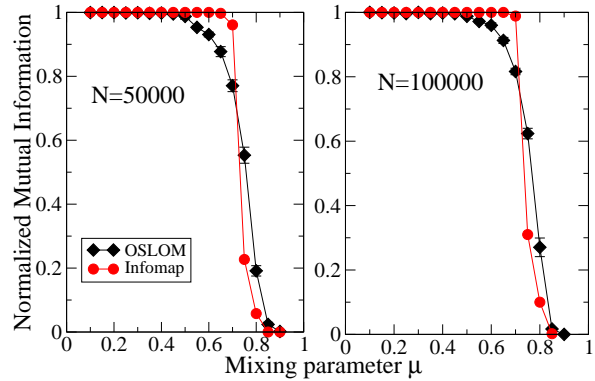}
\caption{ Tests on large undirected and unweighted LFR benchmark graphs
  without overlapping communities. The network sizes are $N=50000$
  (left) and $N=100000$ (right), the maximum degree
$k_{max}=200$ and the community size ranges from $20$ to $1000$.
The other parameters are the same as those used for the graphs of
Fig. 6. 
The two curves refer to OSLOM (diamonds) and Infomap (circles).}
\label{fig6}
\end{center}
\end{figure}
As a term of comparison we used Infomap~\cite{rosvall08}, which has
proved to be very accurate on artificial benchmark graphs~\cite{lancichinetti09c}.
Fig. 6 shows the comparative performance of OSLOM and Infomap
on the LFR benchmark, with undirected and unweighted edges and 
non-overlapping clusters. As a measure of
similarity between the planted partition and that recovered by the
algorithm we adopted the Normalized Mutual Information (NMI)~\cite{danon05},
in the extended version proposed in Ref.~\cite{lancichinetti09},
which enables one to compare both partitions and covers. 
We used this definition also for hard planted partitions,
since modules found by OSLOM may be overlapping.
In all tests on artificial graphs each point is always an average
over $100$ realizations.

The plots correspond to 
two network sizes, $N=1000$ and $N=5000$, and two 
ranges of community size, $[10, 50]$ (``small'') and $[20, 100]$
(``big''), that we indicate with the letters S and B,
respectively. In this way we can check how much the performance of the
algorithm is affected by the network size and the average size of the
communities. The other network parameters are given in the caption. 
From the plots we conclude that OSLOM and Infomap have a basically
equivalent performance.

It is important to test the performance of the algorithms on large
graphs as well, given the increasing availability of large networked
datasets. The question is if and how their performance is affected by
the network size. Fig. 7 shows that both
OSLOM and Infomap are effective at finding communities on large LFR
graphs. We remark that the inferior accuracy of OSLOM when communities are
better defined comes from the fact that the method occasionally finds
homeless vertices, i.e. vertices that are not significantly linked to
any cluster. These are vertices that happen not to have a significant
excess of neighbors
within their community with respect to the number of neighbors in the
other communities, despite the fact that the average number of
internal neighbors is high. This happens because of fluctuations, and the
method judges such vertices as not belonging to any group, which
makes sense. This issue of the homeless vertices is a general feature
of OSLOM. One should not judge it negatively, though. If a vertex $i$
happens to have a number of external neighbors which is appreciably
higher than the expected external degree of the vertex $\mu\,k_{i}$, the condition $p>q$ of the planted
$\ell$-partition model does not hold, so in principle the vertex
should not be put in its original community. The confusion derives
from the fact that the condition $p>q$ holds {\it on average}.

\subsubsection{LFR benchmark with overlapping communities}

The LFR benchmark also accounts for overlapping
communities, by assigning to each vertex an equal number of neighbors
in different clusters~\cite{lancichinetti09b}. 
\begin{figure*}
\begin{center}
\includegraphics[width=3in]{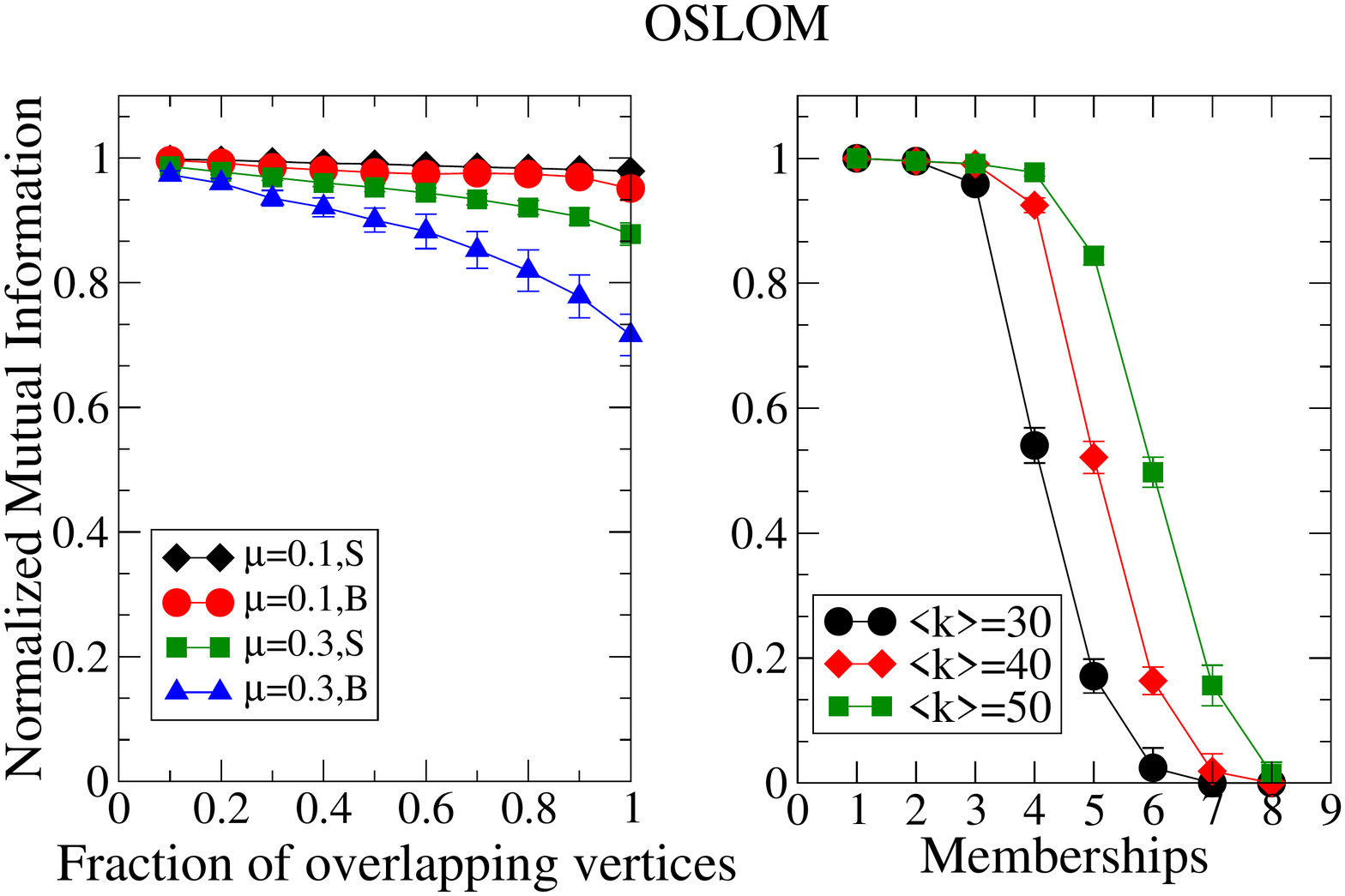}
\includegraphics[width=3in]{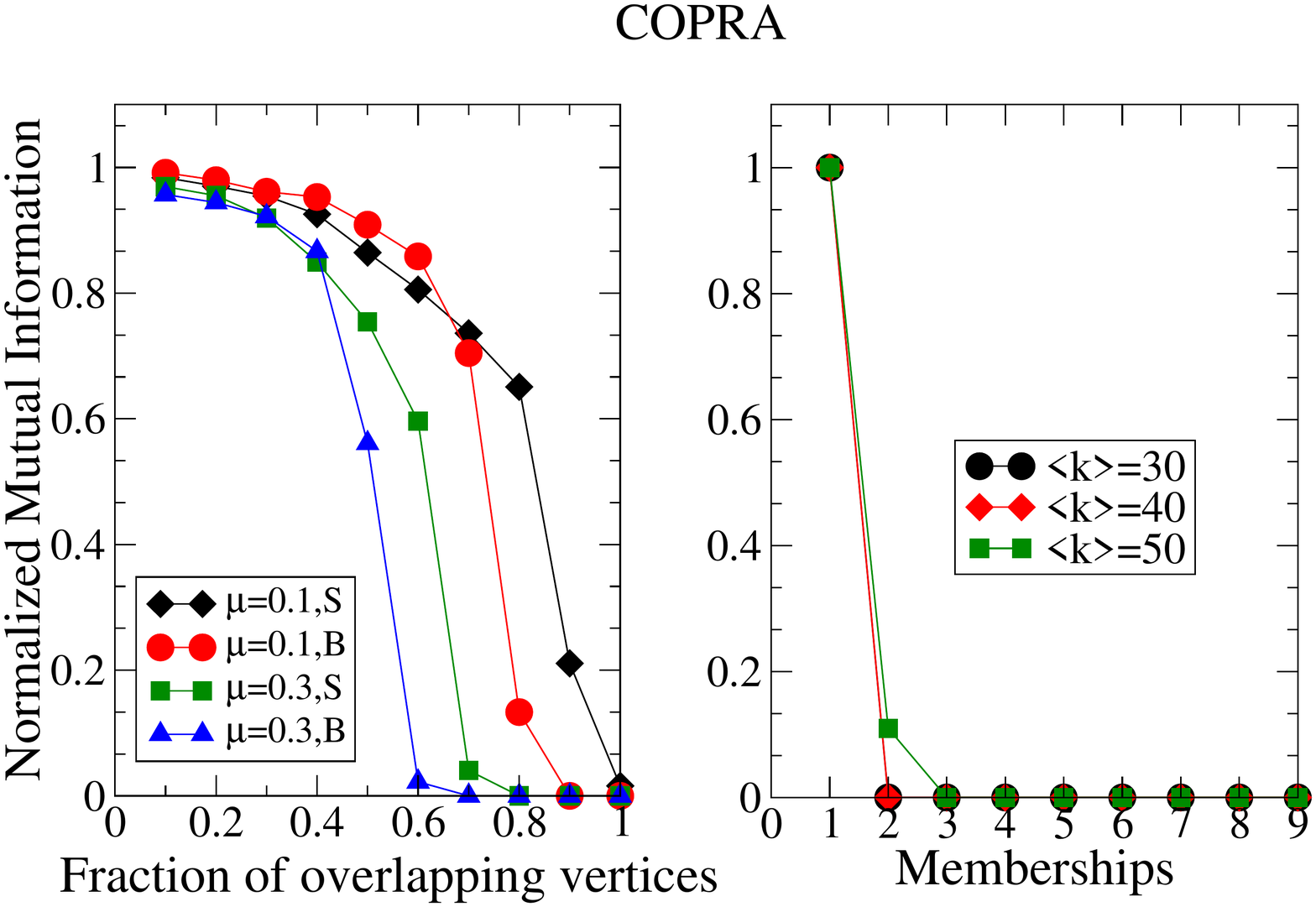}
\includegraphics[width=3in]{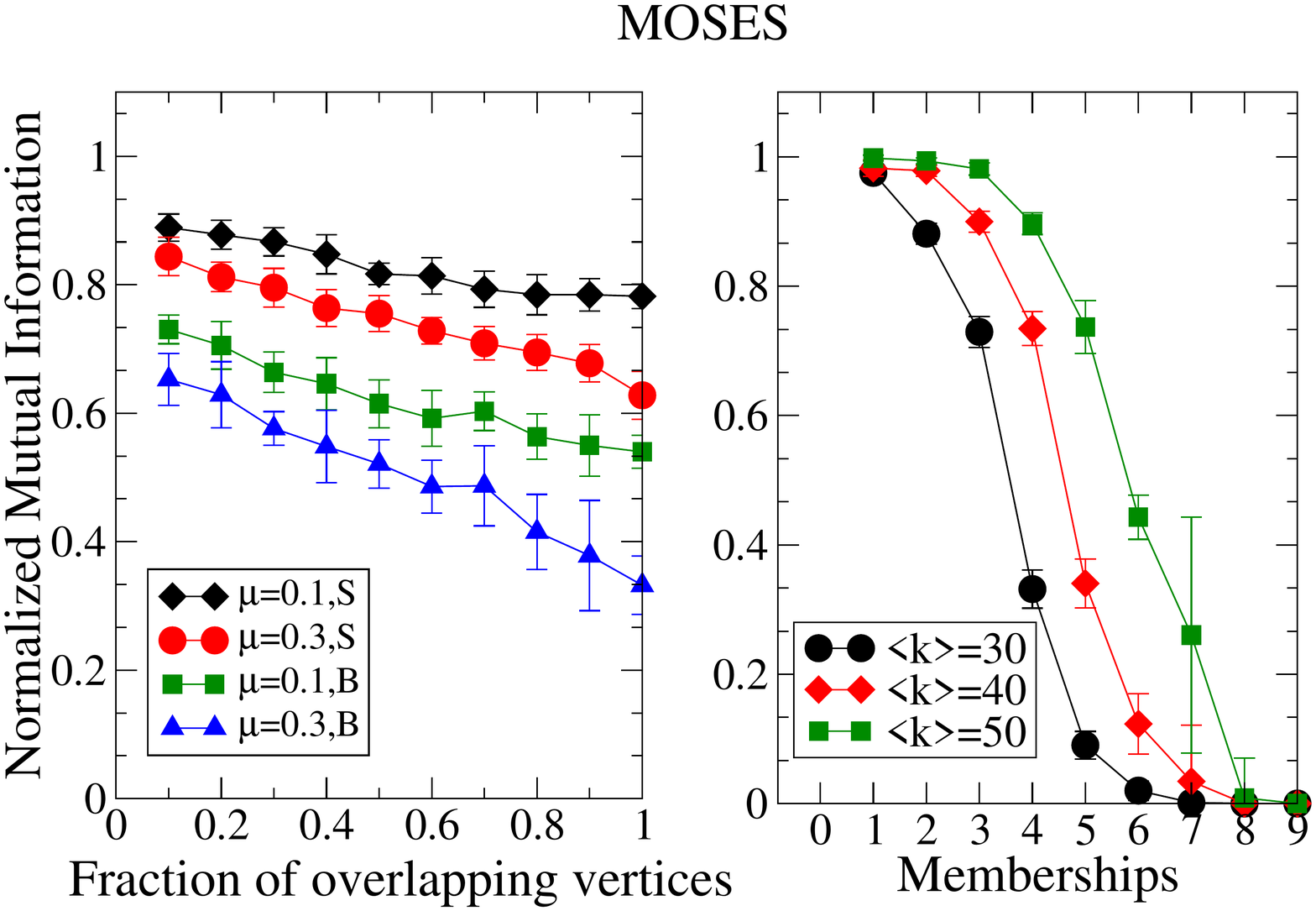}\caption{Test on undirected and unweighted LFR benchmark with
overlapping communities. The parameters are: $N=1000$, $\langle
k\rangle=20$, $k_{max}=50$, $\tau_1=2$,
$\tau_2=1$. S and B indicate the usual ranges of
community sizes we use: $[10, 50]$ and $[20, 100]$, respectively . We tested OSLOM against two
recent methods to find covers in graphs: COPRA~\cite{gregory10} and
MOSES~\cite{mcdaid10}. The left panel displays the normalized mutual
information (NMI) between the planted cover and the one recovered by the
algorithm, as a function of the fraction of overlapping vertices. Each
overlapping vertex is shared between two clusters. The
four curves correspond to different values of the mixing parameter
$\mu$ ($0.1$ and $0.3$) and to the community size ranges S and B. 
The right panel shows a test on graphs whose vertices are all shared
between clusters. Each vertex is member of the same number of
clusters. The plot shows the NMI as a function of the number of
memberships of the vertices. Each curve corresponds to a given value
of the average degree $\langle k\rangle$. The graph parameters are 
$N=2000$, $k_{max}=60$,  $\mu= 0.2$, $\tau_1=2$,
$\tau_2=1$. Community sizes are in the range $[20, 50]$.}
\end{center}
\end{figure*}
To simplify things, we assume
that each vertex belongs to the same number of communities. 
We cannot use Infomap for the comparison, as it delivers ``hard''
partitions, without overlaps between clusters. So we used two recent
methods, that have a good performance on LFR graphs with overlapping
communities: COPRA~\cite{gregory10}, based on label
propagation~\cite{raghavan07}, and MOSES~\cite{mcdaid10}, based on
stochastic block modeling~\cite{nowicki01}. COPRA and MOSES are more
efficient to detect overlapping communities in LFR benchmark
graphs than the popular Clique Percolation Method (CPM)~\cite{palla05},
which is the reason why we do not use the CPM here.
In Fig. 8 we show how the performance of each method decays  
with the fraction of overlapping vertices, for different choices of
the mixing parameter and for the small (S) and big (B) 
communities defined above.
Since in social networks there may be many vertices
belonging to several groups, we also considered the extreme situation
of graphs consisting entirely of overlapping vertices. In this case,
by increasing the number of memberships of the vertices 
communities become more fuzzy and it gets harder and harder for any method to
correctly identify the modules.  From Fig. 8 we deduce that
OSLOM significantly outperforms COPRA in both tests and MOSES in the
test with overlapping and non-overlapping vertices, while the
performances of OSLOM and MOSES are quite close when all vertices are
overlapping.

\subsubsection{Hierarchical LFR benchmark}
\label{test_hier}

OSLOM is capable to handle hierarchical community structure as
well. To test its performance we have designed an algorithm that produces
a version of the LFR benchmark with hierarchy. To keep things simple,
we consider a two-level hierarchical structure (Fig. 9). 
\begin{figure}
\begin{center}
\includegraphics[width=\columnwidth]{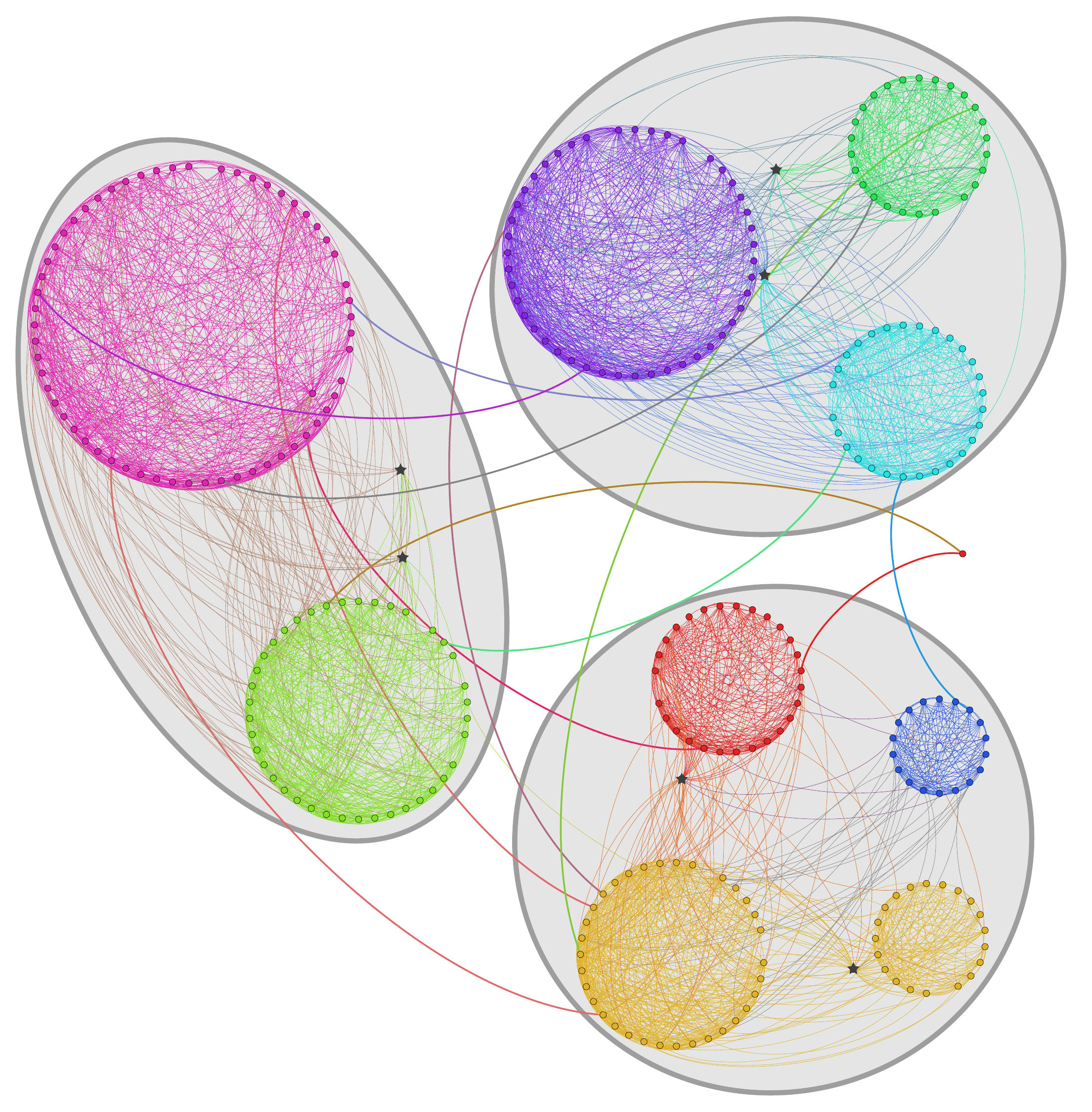}
\caption{A realization of the hierarchical LFR benchmark with two levels. Stars indicate overlapping vertices.}
\end{center}
\end{figure}
The idea is to use the wiring
procedure of the original algorithm twice, first for the
micro-communities and then for the macro-communities.
In order to do so, we need two mixing parameters: $\mu_1$, 
the fraction of neighbors
of each vertex belonging to different macro-communities; $\mu_2$, 
the fraction of
neighbors of each vertex belonging to the same macro-community but to
different micro-communities.
The question is whether the algorithm is able to recover both  
planted partitions of the benchmark, which we call
\textit{Fine} (micro-communities) and \textit{Coarse}
(macro-communities). The partitions found
by the algorithm can be one, two or more, we call 
them partition $1, 2, 3 \dots$. 
\begin{figure}
\begin{center}
\includegraphics[width=3in]{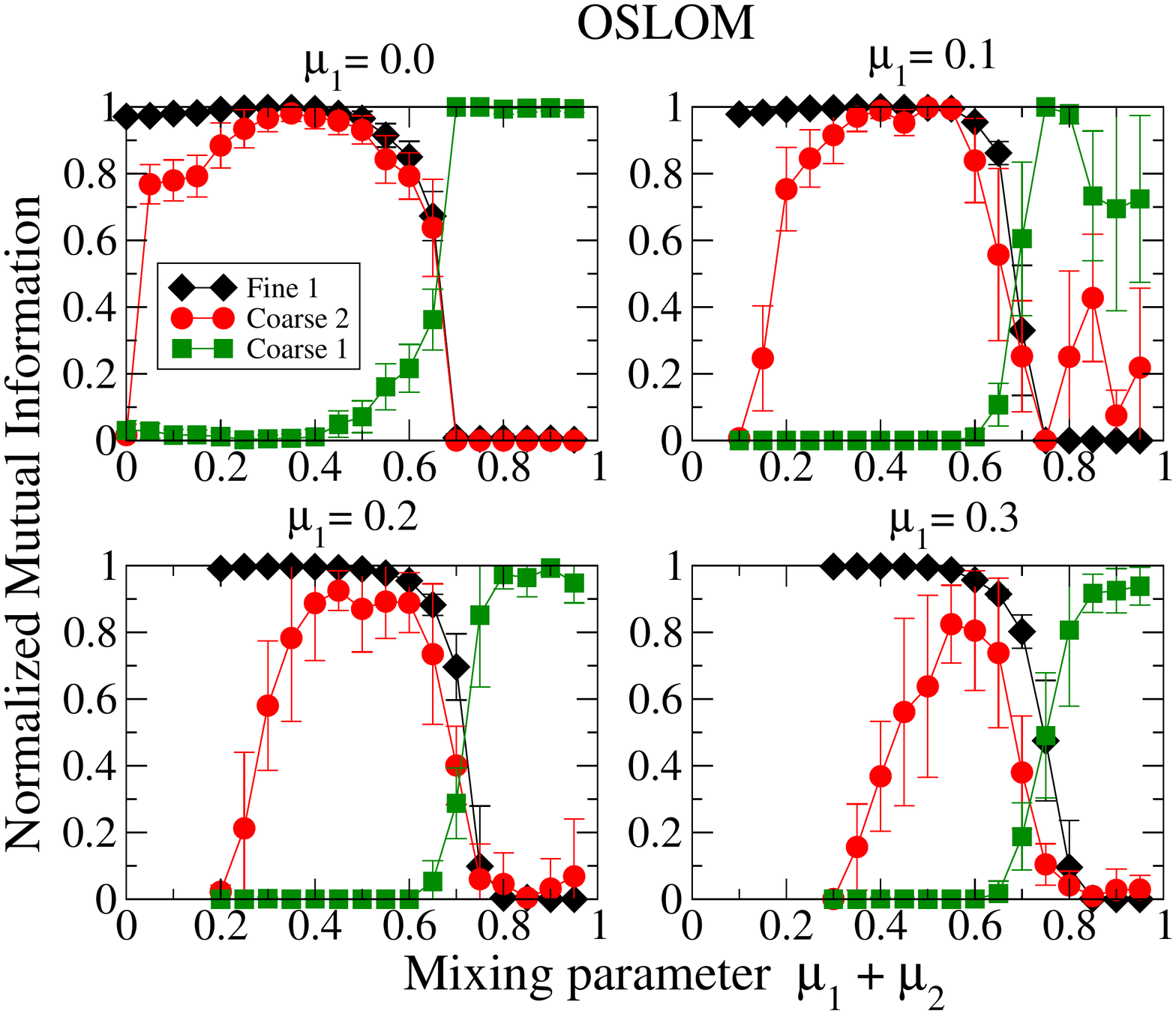}
\includegraphics[width=3in]{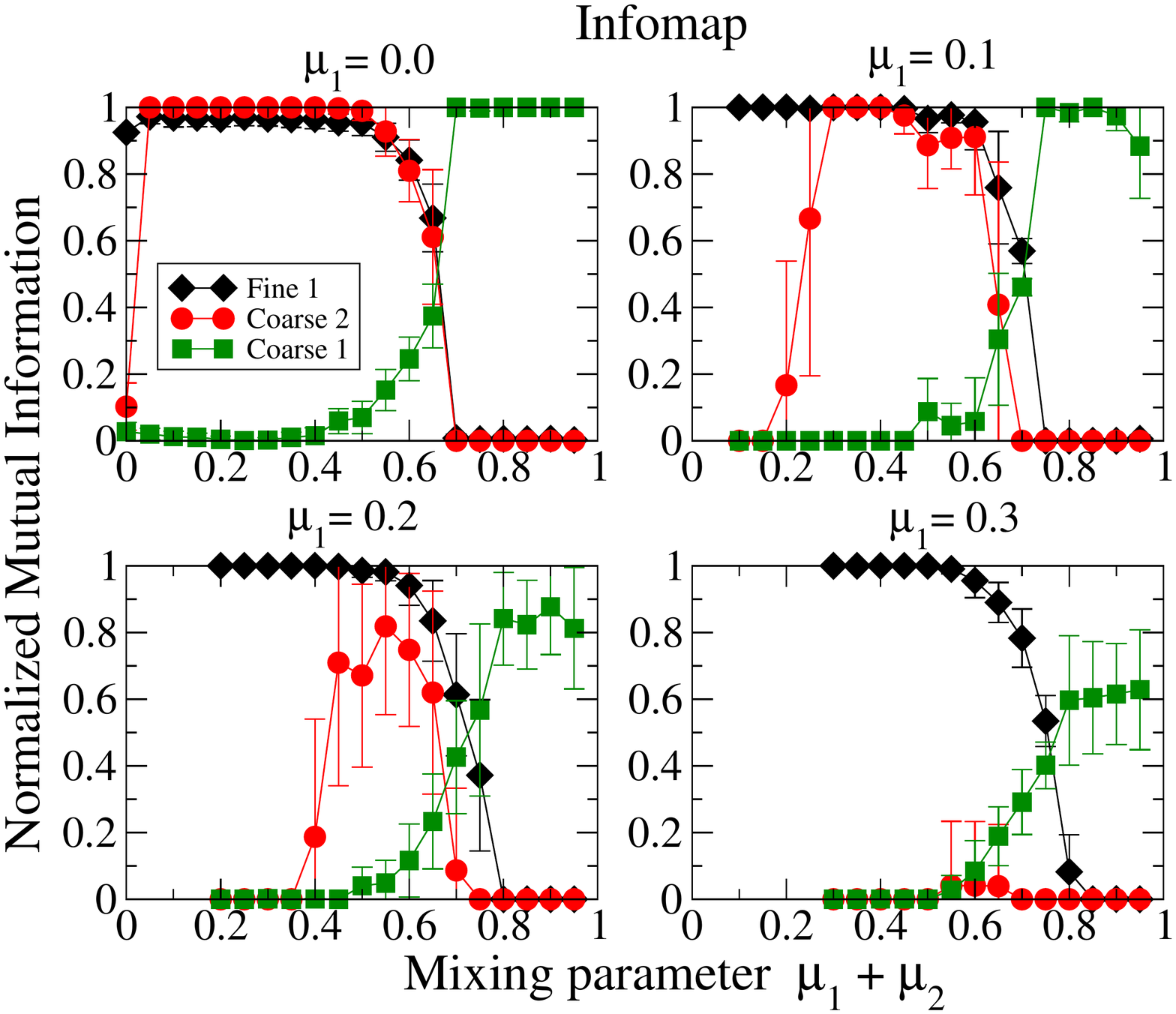}
\caption{Test on hierarchical LFR benchmark graphs (unweighted,
undirected and without overlapping clusters). We compare three pairs
of partitions:
the lowest hierarchical partition found by the algorithm (indicated by
$1$) with the set of micro-communities of the benchmark (Fine); 
the lowest hierarchical partition found by the algorithm with 
the set of macro-communities of the benchmark (Coarse); 
the second lowest hierarchical
partition found by the algorithm (indicated by $2$) with the 
set of macro-communities of the benchmark. The corresponding
similarities are plotted as a function of $\mu_1+\mu_2$, for fixed $\mu_1$.
There are $10000$ vertices, the average degree $\langle k\rangle=20$,
the maximum degree $k_{max}=100$, the size of the macro-communities
lies between $400$ and $4000$ vertices, the size of the
micro-communities lies between $10$ and $100$ vertices. 
The exponents of the degree and community size distributions are
$\tau_1=2$ and $\tau_2=1$.}
\end{center}
\end{figure}
In the test, whose results are illustrated in Fig. 10, 
we compare the Fine partition with partition 1 (Fine 1), the Coarse 
partition with partition 2 (Coarse 2), and the Coarse partition with
partition 1 (Coarse 1). 
We compare OSLOM with a recent extension of Infomap to networks
with hierarchical community structure~\cite{rosvall10}. 
In the plots we show how the similarity of the three pairs of
partitions mentioned above varies by increasing $\mu_2$ 
but keeping $\mu_1$ constant (we picked the values $\mu_1=0$, $0.1$, $0.2$,
$0.3$). For a better comparison of the panels we put on the x-axis
the sum $\mu_1+\mu_2$, representing the fraction of
neighbors of a vertex not belonging to its micro-community.
We find that, when $\mu_2$ increases, the Fine partition becomes difficult
to resolve and, for $\mu_1+\mu_2\gtrsim 0.7$, it cannot be 
found anymore and both algorithms can only find the Coarse partition. 
Instead, for smaller value of $\mu_2$, the algorithms can recover both
levels. OSLOM performs better than Infomap if $\mu_1$ is not too small.

\subsubsection{Random graphs and noise}

We check whether OSLOM is also able to recognize 
the {\it absence}, and not simply the presence, of community
structure. In random graphs vertices are connected to each other at
random, modulo some basic constraints like, e. g., keeping some
prescribed degree distribution or sequence. In this way, there are by
definition no groups of vertices that preferentially link to each
other, so there are no communities. There may be subgraphs with an
internal edge density higher than the average edge density of the
whole network, but they originate from stochastic fluctuations (noise). A good
community finding algorithm should be able to recognize
that such subgraphs are false positives, and discard them.
Here we want to see if OSLOM distinguishes ``order'' from ``noise''.
For this purpose, we carried out two tests.
\begin{figure}
\begin{center}
\includegraphics[width=\columnwidth]{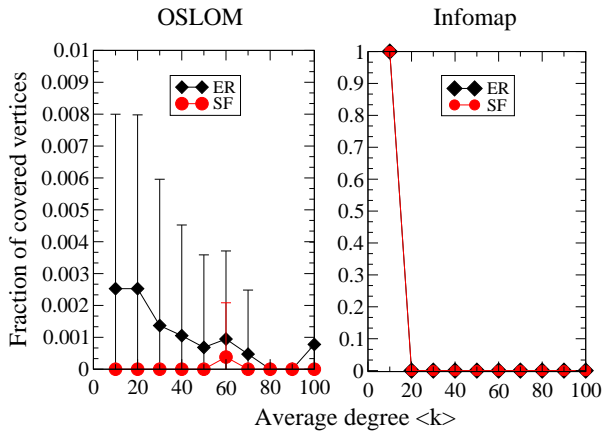}
\caption{Test on random graphs. We plot the fraction of vertices
  belonging to non-trivial clusters (i.e. to clusters with more than
  one and less than $N$ vertices, where $N$ is as usual the size of
  the graph), as a function of the average degree of the graph. The curves correspond to
  Erd\"os-R\'enyi graphs (diamonds) and scale-free networks
  (circles). All graphs have $N=1000$ vertices. The only
  parameter needed to build Erd\"os-R\'enyi graphs is the probability
  that a pair of vertices is connected, which is determined by the
  average degree $\langle k\rangle$. The scale-free
  networks were built with the configuration model [39], starting from a
  fixed degree sequence for the vertices obeying the predefinite power
  law distribution. The parameters of the distribution are: degree exponent $\gamma=2$, maximum degree $k_{max}=200$.}
\end{center}
\end{figure}
In Fig. 11 we applied OSLOM and Infomap to Erd\"os-R\'enyi
random graphs~\cite{erdos59} and scale-free
networks~\cite{barabasi99}. The goal is to see whether the algorithms
recognize that there are no actual communities. Good answers are
the partition with as many communities as vertices, or the partition
with all vertices in the same community. Let us call $\cal P$ the
partition found by the algorithm at hand. Clusters in
$\cal P$ containing at least two vertices and smaller than the whole
network indicate that the method has been fooled. The fraction of
graph vertices belonging to those clusters is a measure of 
reliability: the lower this number, the better the algorithm.  
In Fig. 11 we show this variable as a function of the average
degree $\langle k\rangle$ of the random graphs we considered. 
For OSLOM it remains very low 
for all values of $\langle k\rangle$. This is not surprising, since
OSLOM estimates the statistical significance of clusters, and is
therefore ideal to detect stochastic fluctuations. Infomap instead
finds many non-trivial clusters when $\langle k\rangle$ is low,
whereas it correctly recognizes the absence of community structure if $\langle
k\rangle$ increases.

\begin{figure}
\begin{center}
\includegraphics[width=\columnwidth]{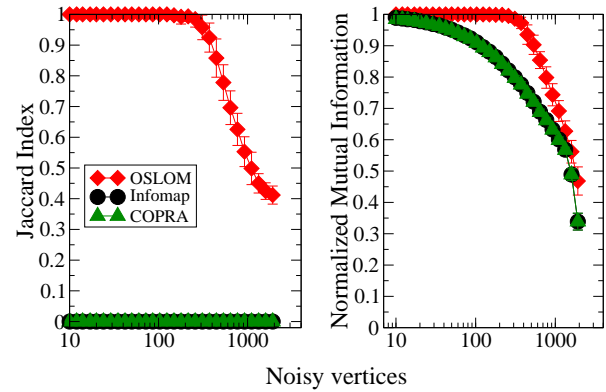}
\caption{Test on graphs including communities and noise. 
The communities are those of an LFR benchmark graph (undirected, unweighted and without overlapping
clusters), with $N =1000$, $\langle k\rangle=20$, $k_{max}= 50$, $\mu=0.2$. The cluster
size ranges from $10$ to $50$ vertices. The noise comes by adding
vertices which are randomly linked to the existing vertices, via
preferential attachment. The test consists in checking whether the
community finding algorithm at study (here OSLOM, Infomap and COPRA)
is able to find the communities of the planted partition of the LFR
benchmark and to recognize as homeless the other vertices.}
\end{center}
\end{figure}

The second test deals with graphs consisting of an {\it
ordered} part, with well-defined clusters, and a {\it noisy} part,
consisting of vertices randomly attached to the rest of the network.
The ordered part is an LFR benchmark graph with $1000$ 
vertices and represents the starting
configuration of our system. The noisy vertices (up to $2000$ in
number) are successively added
in sequence, and a newly added vertex is linked to the other ones
via preferential attachment~\cite{barabasi99}. The initial degree 
of the noisy vertices 
is drawn from a power law distribution with $k_{max}=100$ and exponent
$3$. We measure two things, as a function of the number of noisy vertices: 
the similarity between the set of noisy 
vertices and the set of homeless vertices found by OSLOM, which is
expressed by the Jaccard Index~\cite{tan05} (Fig. 12, left); 
the similarity between
the planted partition of the ordered part of the graph and 
the subset of the partition found by
OSLOM including (only) the vertices of the ordered part, which is
expressed by the normalized mutual information (Fig. 12,
right). We compare OSLOM with Infomap and COPRA~\cite{gregory10}.
We find that OSLOM correctly separates the clusters and the noise up
to a number of about $300$ noisy vertices, which represent almost a
third of the whole network. Infomap and COPRA, instead, do not
recognize the noisy vertices, no matter how small their number
is. Also, they tend to mix noisy vertices with the clusters of
the planted partition of the ordered part, as shown by the fact that
the partition they recover never exactly match the planted
partition, not even when just a few noisy vertices are
present. These results are actually understandable in the case of
Infomap, which is based on the minimization of the code length
required to describe random walks taking place on the graph:
singletons (clusters consisting of single vertices) are generally not
admitted because they increase the amount of information required to map
the process, due to the high number of transitions of the walker from
the singletons to the rest of the graph and back.

\subsection{Real networks}
\label{realn}

\begin{figure*}
\begin{center}
\includegraphics[width=0.8\textwidth] {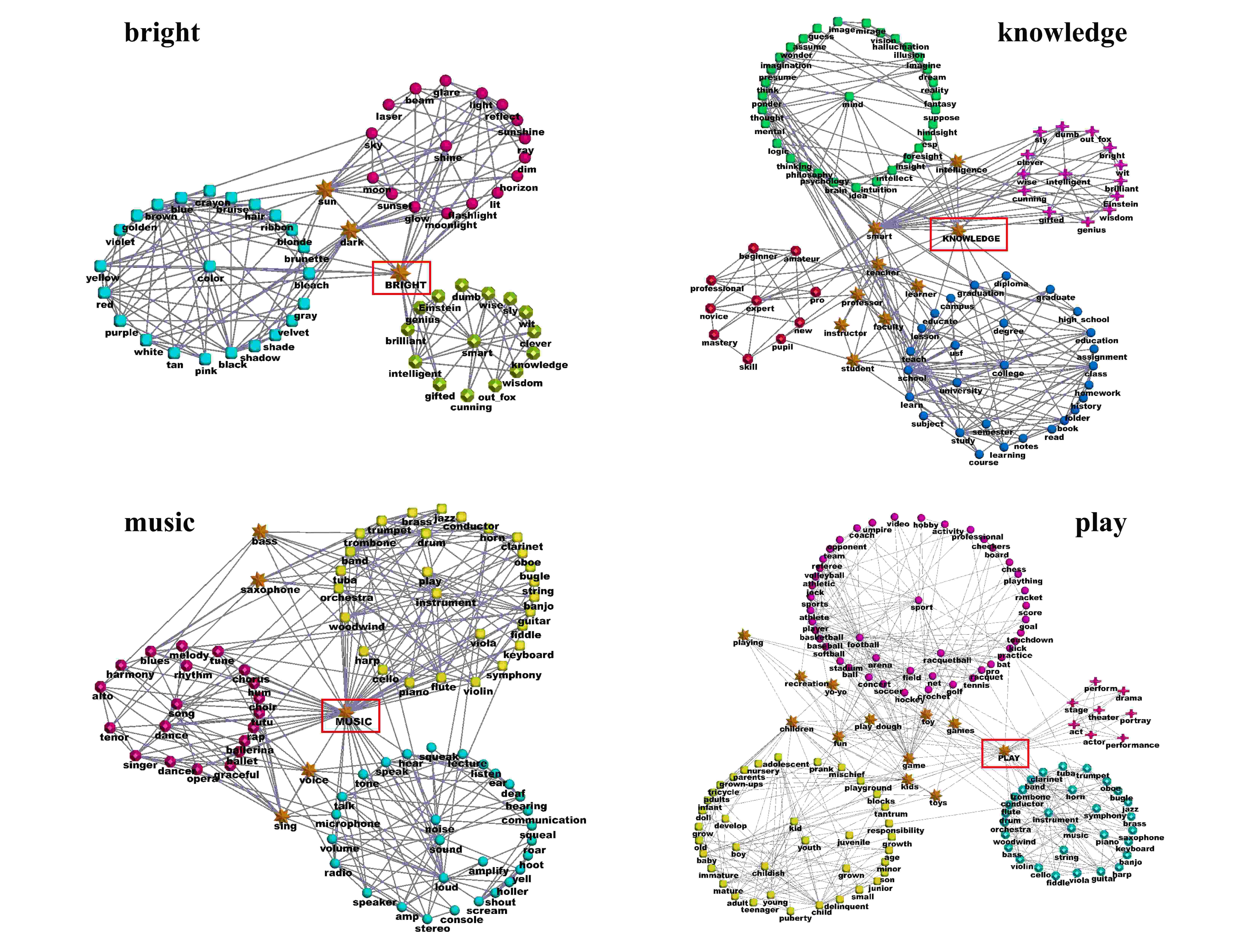}
\caption{Application of OSLOM to real networks: the word association
  network. Stars indicate overlapping vertices.}
\end{center}
\end{figure*}

In this section we discuss the application of OSLOM to networks from the
real world. In Table 1 we list the networks considered
in our analysis, along with some basic statistics obtained from the
detection of their community structure with OSLOM.

\begin{table*}
\begin{tabular}{|c|c|c|c|c|c|c|c|} \hline

Network & N  & E & $\langle k\rangle$  & $N_c$ &
$\langle s \rangle$& $\langle m\rangle$ & $f_h$  \\ \hline
Zachary's club& 34  & 78 & 4.59  & 2 & 17.0 & 1.03 & 0.0294  \\ \hline
Dolphins & 62  & 159 & 5.13  & 2 & 32.5 & 1.08 & 0.0322  \\ \hline
Football & 115  & 613 & 10.7  & 11 & 10.0 & 1.00 & 0.0434 \\ \hline
UK commuting & $10\,608$  & $1\,220\,337$ &  230.07 & 248 & 45.43 & 1.06 &
0.00386  \\ \hline
C. elegans & 453  & $2\,025$ &  8.94 & 25 & 17.04 & 1.22 & 0.229  \\ \hline
Word association & $7\,207$  & $31\,784$ &  8.82 & 261 & 22.48 & 1.35 & 0.395
 \\ \hline
Live Journal & $4\,846\,609$  & $42\,851\,237$ &  17.6 & $407\,451$ & 10.01 &
1.19 & 0.294  \\ \hline
www. uk & $18\,484\,117$  & $292\,244\,462$ &  15.81 & $590\,257$ & 28.08 & 1.02
& 0.125  \\ \hline
US airports 2009 (jan) & 448  & $7\,659$ &  34.19 & 11 & 33.81 & 1.28 &
0.352  \\ \hline
US airports 2009 (mar) & 456  & $8\,491$ &  37.24 & 6 & 67.83 & 1.22 &
0.272  \\ \hline
US airports 2009 (jun) & 453  & $8\,480$ &  37.42 & 9 & 45.33 & 1.28 &
0.315  \\ \hline
US airports 2009 (sep) & 452  & $7\,870$ &  34.81 & 9 & 41.55 & 1.26 &
0.347  \\ \hline
\end{tabular}
\caption{Basic statistics of the real networks we analyzed, including
  the main features of their community structure, detected by OSLOM.
From left to right, we list the number of vertices $N$ and edges
$E$, the average degree $\langle k\rangle$, the number of clusters
$N_c$, the average cluster size $\langle s \rangle$, the average
number of memberships per vertex $\langle m\rangle$ and the fraction
$f_h$ of vertices not assigned to any cluster (homeless vertices). The
values related to the community structure refer to the lowest
hierarchical level.}
\label{table1}
\end{table*}

We analyzed different types of systems:
social, information, biological and infrastructural networks. 
Here we discuss only some of them, the rest of the analysis can be found in Appendix~\ref{real_app}.

\subsubsection{The word association network}

This network is built on the University of South Florida Free
Association Norms~\cite{nelson98}. Here the presence of 
an edge between words $A$ and $B$ indicates that some people associate $B$ to the word $A$.
This network is considered a paradigmatic example of graph with
overlapping communities~\cite{palla05}, since several words may have
various meanings and belong to different groups of words. In Fig. 13 we see
a few subgraphs of the word association network, revolving around four
keywords: {\it bright}, {\it knowledge}, {\it music} and {\it play}. We see that the keywords
are shared among several clusters, which are semantically highly homogeneous.
For instance, {\it bright} belongs to three groups, centered on the
words {\it color}, {\it shine} and {\it smart}, respectively, which
makes sense. In the same subgraph, the words {\it sun} and {\it dark}
are also overlapping vertices, belonging to the groups of {\it color}
and {\it shine}, as one might expect. In the subgraph centered on {\it
  knowledge}, one distinguishes the groups referring to the words
{\it mind}, {\it intelligent}, {\it expert} and {\it college}/{\it
  university}. Here there are many overlapping vertices, like the word
{\it intelligence}, shared between the groups of {\it mind} and {\it
  intelligent}, and a bunch of terms indicating (mostly) professional status
within schools and/or universities, like {\it student}, {\it
  professor}, {\it teacher}, etc., which lie between the groups of
{\it expert} and {\it college}/{\it university}. In the third
subgraph, the word {\it music} is shared by the groups of {\it
  instrument}, {\it song}/{\it dance} and {\it noise}/{\it sound}:
other overlapping vertices are the words {\it sing} and {\it voice},
lying between {\it song}/{\it dance} and {\it noise}/{\it sound}, and
the words {\it bass} and {\it saxophone}, belonging to the groups
of {\it song}/{\it dance} and {\it instrument}. Finally, the word {\it
  play} sits between the communities of {\it sport}, {\it music} and
{\it youth}/{\it kid}; other overlapping vertices in this subgraph
include {\it game}, {\it children}, {\it toy}, etc..

\subsubsection{UK commuting}

\begin{figure*}[t]
\begin{center}
\includegraphics[width=\textwidth] {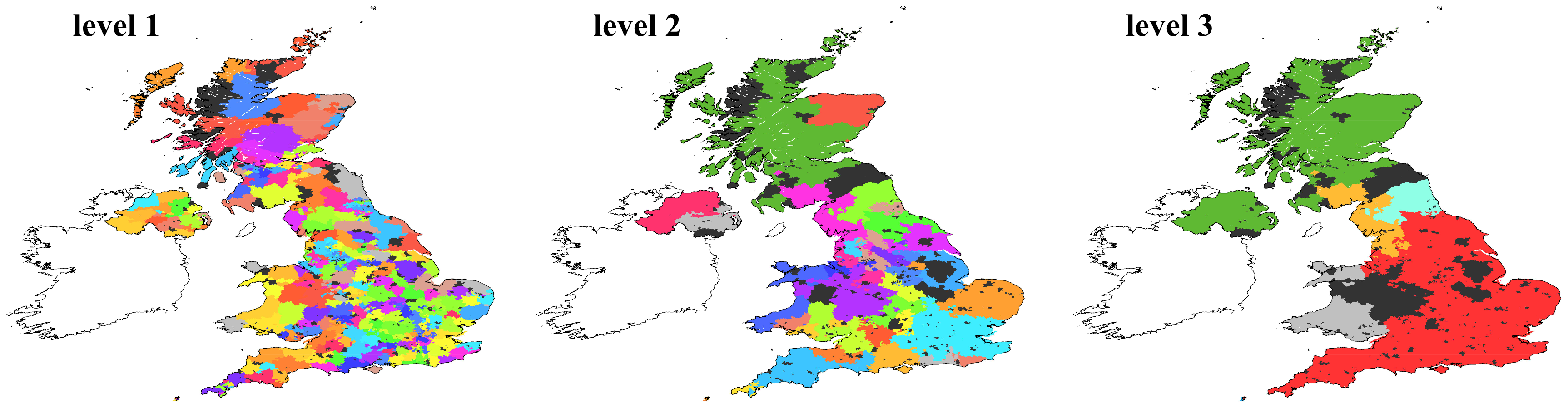}
\caption{Application of OSLOM to real networks: flows of commuters in
  the UK. Black points indicate overlapping vertices.}
\end{center}
\end{figure*}

This is the network of flows of commuters between areas of the United
Kingdom, and therefore it has a clearly geographic character.
It is composed of $10\, 608$ vertices, each representing a
ward, i. e. 
a geographical division used in the UK census for statistical purposes. The whole
territory of the United Kingdom is divided into wards. 
Each edge corresponds to a flow of
commuters between the ward of origin and that of destination, with a weight
accounting for the number of commuters per day. The data were
collected during the
$2001$ UK census, when the ward of residence and the ward of work/study was
registered for a sizeable part of the British population. The database can
be accessed online at the site of the Office for National Statistics
{\tt http://www.ons.gov.uk/census}. OSLOM finds three hierarchical
levels (Fig. 14). The clusters of the second level delimit
geographical areas typically centered about one major town. In the highest level the areas of England,
Wales, Scotland and Northern Ireland are clearly recognizable. Interestingly, Northern
Ireland and Scotland are parts of the same community, due to the large
flow of commuters between the two regions, despite the geographical separation. Black points represent
overlapping vertices.

\subsubsection{LiveJournal and UK Web}

We also applied OSLOM to two large networks. The first is  
a network of friendship relationships between users of the
on-line community {\it LiveJournal} ({\tt www.livejournal.com}), and
was downloaded from the Stanford Large Network
Dataset Collection ({\tt http://snap.stanford.edu/data/}). The second
is a crawl of the Web graph carried out by the Stanford WebBase
Project ({\tt http://dbpubs.stanford.edu:8091/}
{\tt $\sim$testbed/doc2/WebBase/}), within the
UK domain ({\tt .uk}). 
\begin{figure}
\begin{center}
\includegraphics[width=\columnwidth] {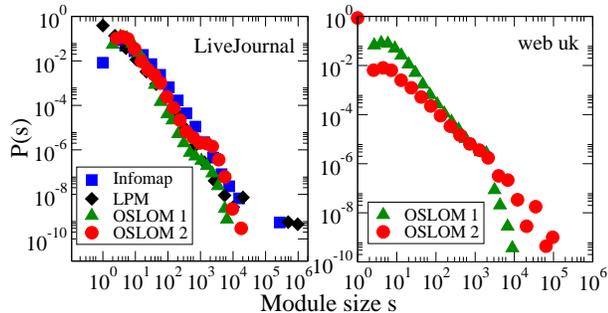}
\caption{Application of OSLOM to real networks: friendships of {\it
  LiveJournal} users (left) and sample of the {\tt .uk} domain of
  the Web graph (right). We show the distribution of cluster sizes
  obtained by OSLOM for the first two hierarchical levels (OSLOM 1 and
OSLOM 2). For {\it
  LiveJournal} we can compare the distributions with those found with
Infomap~\cite{rosvall08} 
and the Label Propagation Method (LPM) by Leung et al.~\cite{leung09}.  }
\end{center}
\end{figure}
We remind that the Web graph is a directed graph whose vertices are
Web pages, while the edges are the hyperlinks that enable one to surf
from one page to another. 
These two systems are too large for OSLOM, due to the huge variety of
possible cluster sizes to explore. Therefore we applied a two-step method: in the first step, we derived
an initial partition $\cal P^\star$ with the Louvain method~\cite{blondel08}, which
is able to handle large networked datasets; in the second step, we apply OSLOM to
refine the clusters of $\cal P^\star$. In principle, this procedure should yield the
same partitions/covers as applying OSLOM directly, if one repeated
OSLOM's cluster search many times. But this would make the
calculations too lengthy, so, in order to complete the analysis within a
reasonable time, it is necessary to keep the number of iterations low.  
In this way there is the big advantage of
drastically reducing the computational complexity, which makes large
systems tractable, even if results would be more accurate if one could
apply OSLOM from scratch. Clearly, since different iterations are
independent processes, one could sensibly increase the statistics by
distributing the iterations among different processors, if
available. 
\begin{figure*}
\begin{center}
\includegraphics[width=\textwidth] {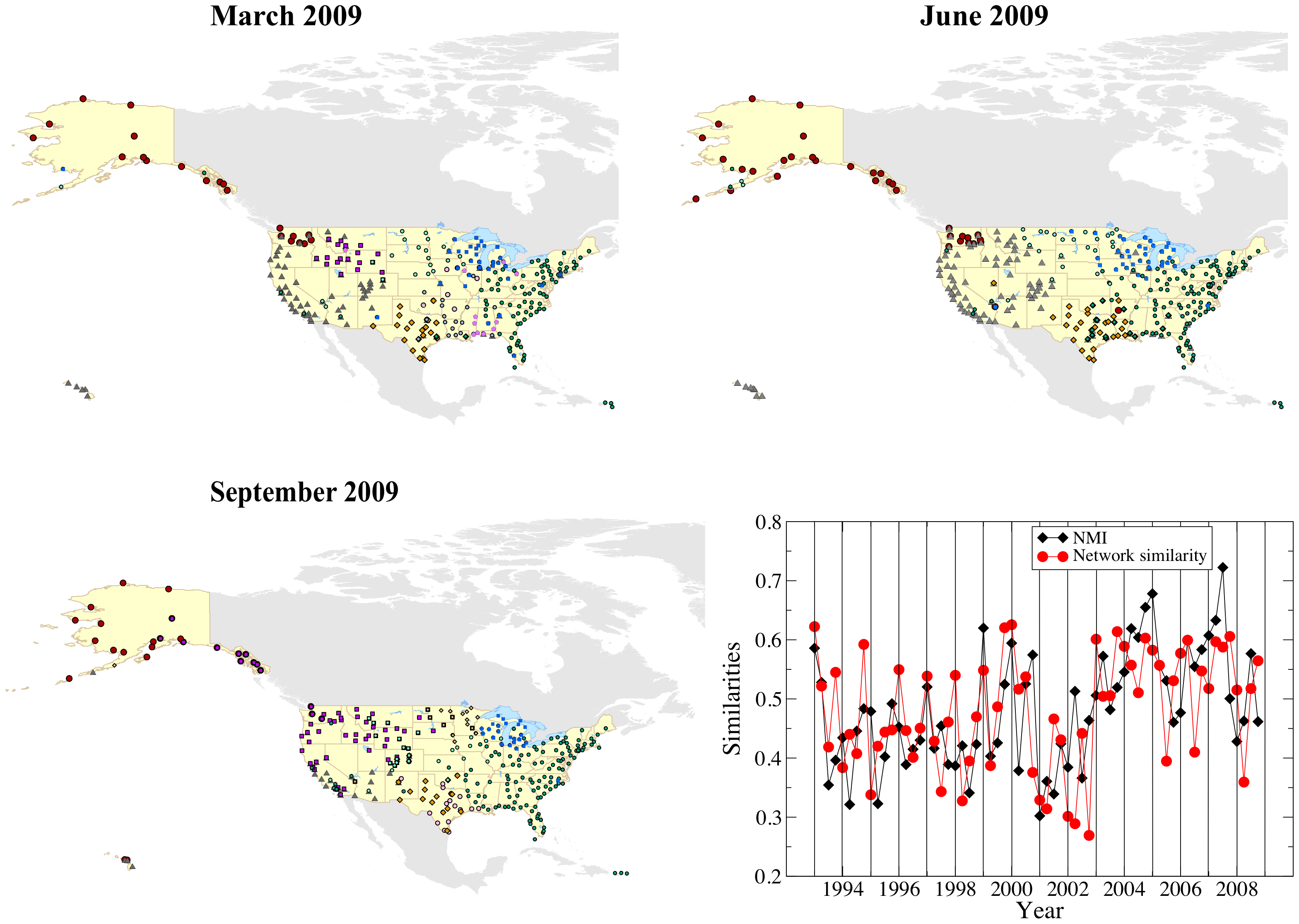}
\caption{Application of OSLOM to real networks: US airport
  network. The maps show the position of the airports, which are
  represented by symbols, indicating the communities found by applying
  OSLOM directly to the corresponding network, without exploiting the
  information of previous snapshots. The diagram shows the
  ``seasonality'' of air traffic. The
  normalized mutual information (diamonds) was computed comparing the cover of the system at time
$t$ adjusted by OSLOM on the network at time $t+\Delta t$, and the
cover obtained by applying OSLOM directly to the system at
time $t+\Delta t$. The circles are estimates of the
similarity of the network matrices of snapshots separated by $\Delta
t$ (one year). For each year we took four snapshots, by cumulating the traffic
of each trimester. The most stable networks are typically in winter (vertical lines).}
\end{center}
\end{figure*}

In Fig. 15 we present the
distribution of cluster sizes of the first two hierarchical levels
found by OSLOM. The results are obtained by performing a single
iteration on a workstation HP Z800. For the Web graph, which is the
larger system, with nearly $20$ million vertices and $300$ million
edges (see Table 1), the analysis was completed in about $40$ hours.
For the social network of {\it LiveJournal} we can
compare the results with the corresponding distributions found by Infomap and the Label
Propagation Method (LPM) proposed by Leung et al.~\cite{leung09},
which were computed in a recent
analysis~\cite{lancichinetti10c}. In that work the original Infomap was used,
so neither Infomap nor the LPM could detect hierarchical community
structure and there is just one cluster size distribution,
corresponding to the single partition recovered. The distributions are
broad and quite similar across different methods. Interestingly, the two
hierarchical levels of {\it LiveJournal} (OSLOM 1 and OSLOM 2) are
not too different, indicating a sort of self-similarity of the
community structure. For the Web the two levels are more dissimilar and
the distributions have a clear power law decay (with different
exponents) up to a cutoff, which is approximately the same for both
curves ($\sim 2000$ vertices). 

\subsubsection{Dynamic datasets: the US air transportation network}

For the last application, we used a time-stamped dataset, the US air
transportation network. The data
can be downloaded from the
Bureau of Transportation Statistics (US government) ({\tt http://www.bts.gov}).
Vertices are airports in the USA and edges are weighted by the number of
passengers transported along the corresponding routes. In
Fig. 16 we show the geographical location of the airports and
their communities, indicated by the symbols, for three snapshots,
corresponding to the traffic in March, June and September 2009,
respectively. We remind that for dynamical datasets we usually take the
partition/cover ${\cal P}(t)$ of the system at time $t$, and we use it as initial
partition/cover for the topology of the system at time $t+\Delta t$,
which is then refined by OSLOM, in order to ``adapt'' ${\cal P}(t)$ to the
current structure. This is done to exploit the information of more
snapshots at the same time. Since the three maps of Fig. 16 are mostly
illustrative, communities were derived by applying directly OSLOM to the
corresponding snapshots, for simplicity.
The diagram indicates the similarity between
networks and their corresponding partitions/covers in different
snapshots. 
Each snapshot represents the whole traffic of one trimester, which
corresponds to a season, while $\Delta t=1$ year, as we want to
measure the variation of the network structure in consecutive seasons.  
The similarity between partitions/covers is computed with
the normalized mutual information, as usual.
The similarity of two weighted networks like the ones at study is
measured in the following way. First, one computes the distance
$d_{t,t+\Delta t}$ between the matrices ${\bf {\tilde{W}}}^{t}$ and
${\bf {\tilde{W}}}^{t+\Delta t}$:  $d_{t,t+\Delta
  t}=\sqrt{\sum_{ij}({\tilde{W}}_{ij}^t-{\tilde{W}}_{ij}^{t+\Delta t})^2}$. The
  matrix ${\bf {\tilde{W}}}^{t}$ is derived from the standard weight matrix
  ${\bf {W}}^{t}$ by dividing each edge weight by the sum of all edge
  weights. This is done because the traffic flows tend to increase
  steadily in time, so comparing the original weight matrices is not appropriate.
The quantity $d_{t,t+\Delta t}$ is a dissimilarity measure. We turn it
to a similarity index by changing its sign, adding a
constant and rescaling the resulting values. Since we wish to compare the trend of the network similarity
with that of the partition/cover similarity, the additional constant and the
rescaling factor are chosen such to reproduce the average and the variance
of the curve of the normalized mutual
information. After this operation, the two trends are finally comparable.
The diagram shows that both measures follow a yearly periodicity, with
peaks corresponding to the winter season, which is then more stable
than the others.

\section{Discussion}

We have introduced OSLOM, the first method that finds clusters in networks
based on their statistical significance. It is a
multi-purpose technique, capable to handle various types of graphs,
accounting for edge direction, edge weights, overlapping communities, hierarchy
and network dynamics. Therefore, it can be used for a wide variety
of datasets and applications.

We have thoroughly tested OSLOM against the
best algorithms currently available on various types of artificial
benchmark graphs, with excellent results. In particular, OSLOM is
superior on directed graphs and in the detection of strongly
overlapping clusters. Moreover, it is an ideal method to recognize 
the absence of community structure and/or the
presence of randomness in graphs. In some cases 
OSLOM returns slightly less accurate results than other methods, 
because it finds several homeless vertices when
communities are fuzzy. This is due to the fact that, in the
realizations of benchmark graphs, it may happen that some vertices
end up having the same number of neighbors (or even more) in other communities than in their own,
due to fluctuations, even if on average this does not happen. So, the classification of those
vertices, {\it imposed} by the planted $\ell$-partition model, is not
justified topologically. This is an important general issue that needs
to be assessed in the future, to avoid systematic errors in the
testing procedure.

OSLOM is a local algorithm, so it respects the nature of community
structure, which is a local feature of
networks, the more so the larger the systems at study. 
However, the null model adopted to estimate the statistical
significance of clusters is the configuration model, which is global. 
This is the same null model adopted in modularity
optimization~\cite{newman06}, and is responsible for the serious
problems of this technique, like its well known resolution
limit~\cite{fortunato07}. Therefore we perform an iterative cluster
search within the clusters found after the first application of the
method, by considering each cluster as a network on its own. 
In this way we progressively limit the horizon of the part of
the network under exploration, and we are able to find the smallest
significant clusters, which are the natural building blocks of the
network and the basis of its hierarchical community structure. So 
the null model, originally global, gets confined to smaller
and smaller portions of the graph. The actual resolution of the method
is thus not due to the null model, but to the choice of the threshold
$P$. In this paper we have set $P=0.1$, which is often used in various
contexts and delivers an excellent performance on the benchmark graphs
we have adopted. Nevertheless, how much a real graph {\it deviates}
from a random graph depends on the specific system at hand, and it
would be more appropriate to estimate the threshold $P$ case by case. This
is an issue to consider for future work. We remark that also for
modularity optimization one could in principle iteratively restrict the null model
to the clusters found by the method. However, modularity is based on
the {\it expected} value of variables estimated on the null model,  
neglecting random fluctuations, which is why modularity can attain large
values on specific partitions of random
graphs~\cite{guimera04,reichardt06b,reichardt07}.
OSLOM instead accounts for those fluctuations, so it is far more
reliable, in this respect. Furthermore OSLOM is a local method, so it
does not suffer from the severe problems coming from modularity's global optimization~\cite{good10}.

Another important aspect to emphasize is the need to perform many
iterations, to get more accurate results. This is not a
specific feature of OSLOM, but it should be done for all community
detection techniques with a stochastic character, like methods based
on optimization (e. g., modularity optimization). In the literature 
there is the general attitude to perform a single iteration, and to
reduce the complexity of an algorithm to the time required to carry
out one iteration. But this is not appropriate, especially on large
networks. For instance, by performing a single iteration, vertices lying on the border between clusters may be
assigned to a specific cluster, while in many cases they are
overlapping. By combining the results of several iterations, instead,
it is more likely to distinguish overlapping vertices from the
others. Furthermore, one can compute the strength of the membership of
vertices in different clusters, from the frequency with which they
were classified in each cluster. One can also disambiguate stable
from unstable clusters, which could be recovered from specific iterations.
So, it is crucial to collect and combine the results of many
iterations. Of course, the complexity of the method grows with the
number of iterations, but it can be considerably reduced by distributing runs among
many different processors, if large computer clusters are
available. 

The running time of OSLOM is dominated by the exhaustive search of
significant vertices, inside and outside the clusters. This search
could be carried out with greedy approaches, with a huge
computational advantage, and this is an improvement we plan to
implement in the near future. On the other hand, if one wishes to attack
very large graphs, OSLOM could be used at a second stage, as a
refinement technique, to clean the results of an initial partition
delivered by a fast algorithm. In this case, since the initial
clusters are usually cores or parts of the significant clusters we are
looking for, OSLOM converges far more rapidly than its direct
application without inputs. 
We have seen in the previous section
that, by combining OSLOM with the Louvain method by Blondel et al., we
were able to handle systems with millions of vertices.

We have proposed a recipe to deal with the increasingly more important
issue of detecting communities in dynamic networks. The idea is to
take advantage of the information of different snapshots at the same
time, by ``adapting'' the partition/cover of the earlier snapshot to
the topology of the other one. In this way it is possible to uncover
the correlation between the structures of the system at different time
stamps. 

We have shown the versatility of OSLOM by applying it to various networked datasets. 
OSLOM provides the first comprehensive toolbox for the analysis of community
structure in graphs and is an ideal complement of existing tools for
network analysis. The algorithm, with all its variants (including a
fast two-step procedure for the analysis of very large networks) is
implemented in a freely downloadable and documented software ({\tt http://www.oslom.org}).

\begin{acknowledgments}

We thank Paolo Bajardi, Steve Gregory and Martin Rosvall for useful suggestions.
A. L. and S. F. gratefully acknowledge ICTeCollective. The project
ICTeCollective acknowledges the financial support of the Future and Emerging Technologies (FET)
programme within the Seventh Framework Programme for Research of the 
European Commission, under FET-Open grant number 238597.

\end{acknowledgments}

\begin{appendix}

\section{Numerical estimation of the internal connection probability}
\label{a1}

The assessment of a cluster's significance given the null
(configuration) model relies on the estimation of the 
probability described in Eq.~\ref{hyper_eq1}. This function
has to be evaluated many times along the execution of OSLOM in order to clean up each
cluster and to evaluate the 
clusters at the different hierarchical levels. We explain here how the
values of the 
distribution function can be estimated or approximated in a practical
implementation of OSLOM.

For convenience, we rewrite the equation here
\begin{equation}
\label{hyper_eq1_app}
p(k_i^{in}|i,\mc,{\mathcal G})= A \frac{2^{-k_i^{in}}}{k_i^{out}! \; k_i^{in}!\; (m_{\mc}^{out}- k_i^{in})! \; (M^*/2)!}.
\end{equation}
While estimating the value of the probability of Eq.~\ref{hyper_eq1_app} for a
certain $k_i^{in}$, the most 
computationally expensive part is the evaluation of the normalization
factor $A$. In fact, 
this would force us to evaluate the rest of the formula for all the
allowed values of $k_i^{in}$ and add up the result. A simple way out of this problem
is to approximate the distribution by another whose 
normalization factor is known. To do so, we can think of a slightly
different null model, in which the edges are still drawn at random and the formation 
of self-loops is admitted. This is actually the null model on which
the definition of modularity is based~\cite{newman04b}. In such model, 
the equivalent of Eq.~\ref{hyper_eq1_app} becomes an hypergeometric function that 
is much easier to estimate (see ~\cite{lancichinetti10}). Both
distributions, that of Eq.~\ref{hyper_eq1_app} and the hypergeometric, provide close numerical
values for the same $k_i^{in}$, except if the probability of generating self-loops
in the null model is high. The probability that reshuffling
the connections at random a stub of vertex $i$ connects to another stub of 
the same vertex, is given by $k_i^2/2M$. In the software implementation of OSLOM, the hypergeometric
approximation for Eq.~\ref{hyper_eq1_app} is used as long as
$k_i^2/2M<1$. Otherwise, 
we directly measure $A$ from Eq.~\ref{hyper_eq1_app}.

\section{Extension of the method to weighted networks}
\label{wnetworks_app}

In the main text, it is briefly discussed how to extend OSLOM to
weighted graphs. 
We mention also that some of the technical issues, such as combining
both $r_w$ and $r_t$, are not trivial. This procedure is described here in further detail.

Remember that we start from an ansatz for the distribution of the
weights in the null model. 
The distribution of the probability of having a certain weight on the 
edge joining vertices $i$ and $j$ was assumed to be
\begin{equation}
\label{p_w_ij_1}
p(w_{ij} > x |k_i,k_j,s_i,s_j)= \exp( - x / \langle w_{ij} \rangle).
\end{equation}
The idea behind this expression is that the weight of an edge is
proportional to the average weight of its endvertices ($\langle w_{i} = \rangle s_i/k_i$ and $\langle w_{j} \rangle = s_j/k_j$).
We proposed the harmonic average because it is more sensitive to small values of $\langle w_{i} \rangle$.
Our goal is to define a fitness function $r$ which has 
to be a uniform random variable on our randomized weighted network.
And we want to combine the fitness function depending on the topology with
one depending on the weight distribution in order to detect meaningful fluctuations in any of them.

Let us consider a vertex $i$ which has $l$ connections with a given
subgraph $\mc$ (not including $i$).
For the topological part, we have already computed the probability
that $i$ shares $l$ or more edges with vertices of $\mc$ (Eq.~\ref{hyper_eq1_app}). We call this number $r_t$.
Each of the $l$ edges joining $i$ with $\mc$ carries a weight. We consider the
corresponding normalized weight
$\omega_s= w_s/ \langle w_{s} \rangle$, where $w_s$ is the weight on
the $s$-th edge, with $s = 1, 2,  \dots, l$. Since we want a single number 
taking into account all the weights in the set, we can simply consider the sum of all the $\omega_s$:
\begin{equation}
\Omega = \sum_{s=1}^l \omega_s
\end{equation} 
$\Omega$ is the sum of $l$ exponentially distributed variables (with
rate equal to one) and therefore it follows the Erlang distribution~\cite{evans00}. 
Let us call $r_w$ the cumulative of $\Omega$:
\begin{equation}
r_w= p(\Omega>x)=  e^{-x}  \sum_{q=0}^{l-1}x^q / q!
\end{equation} 
In this way, we managed to
define two variables $r_t$ and $r_w$ which are both
uniformly distributed in the null model. 
Now, we would like to combine these two scores to have a final
score for our vertex $i$. Unfortunately this is not so simple.
We remind that $r_w$ is defined only on the $N_n$ neighbors of subgraph
$\mc$ while $r_t$ is defined for all the $N^* = N - n_\mc \geq N_n$ vertices out of $\mc$, so the two
variables are defined on samples of different size, in general. A way to overcome this difficulty
is to scale $r_t$ to an equivalent random variable ${r^\prime}_t$ defined on
a smaller sample. 
This amounts to map each index $i$ in the set $1,
2,..., N^*$ of the old variable onto an index $j$ in the set $1, 2,
..., N_n$ of the new variable. Given $i$,
the natural solution is to pick the index $j$ such that the cumulative
probability $\Omega^t_q$ on the sample of $N^*$ vertices  coincides (at
least with the approximation allowed by the specific numerics
involved) with the cumulative probability $\Omega^w_q$ on the smaller
sample of $N_n$ vertices. It can be shown that
this can be achieved with a good approximation (in the limit of $j$
close to $N_n$) with the following rescaling:
\begin{equation}
{r'}_t= r_t \cdot \frac{N^*+1}{ N_n +1}.
\end{equation}
Once we computed ${r'}_t$ and $r_w$ we need to combine them in
order to have a single score to rank the vertices. We 
consider the product ${r'}_t\cdot r_w$ and the final score $r_{tw}=
p( {r'}_t \cdot r_w < x)= x (1 - \log x)$. The last expression comes
from the assumption that the two variables are both uniform and independent. The set of 
variables $\{r_{tw}\}$ is then used to rank the vertices and to
compute the cumulative probabilities $\Omega^{tw}_q$, with
$N_n$ instead of $N^*$.

\section{Further tests on benchmark graphs}
\label{tests_app}

\subsection{Girvan-Newman benchmark}
\label{GN_app}

\begin{figure}
\begin{center}
\includegraphics[width=\columnwidth]{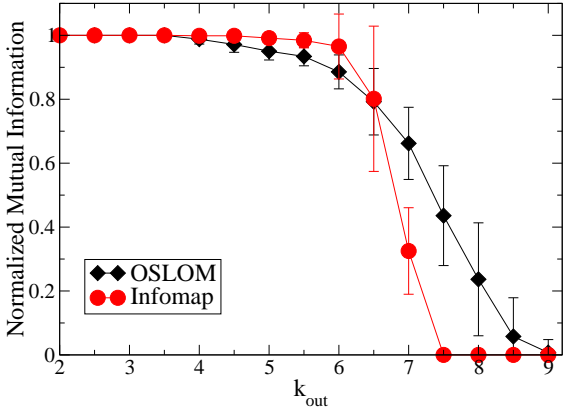}
\caption{Test on the Girvan-Newman benchmark graphs. The variable
  $k_{out}$ is the average number of external neighbors per vertex.  
The two curves refer to OSLOM (diamonds) and Infomap (circles).}
\label{fig4}
\end{center}
\end{figure}
\begin{figure}
\begin{center}
\includegraphics[width=\columnwidth]{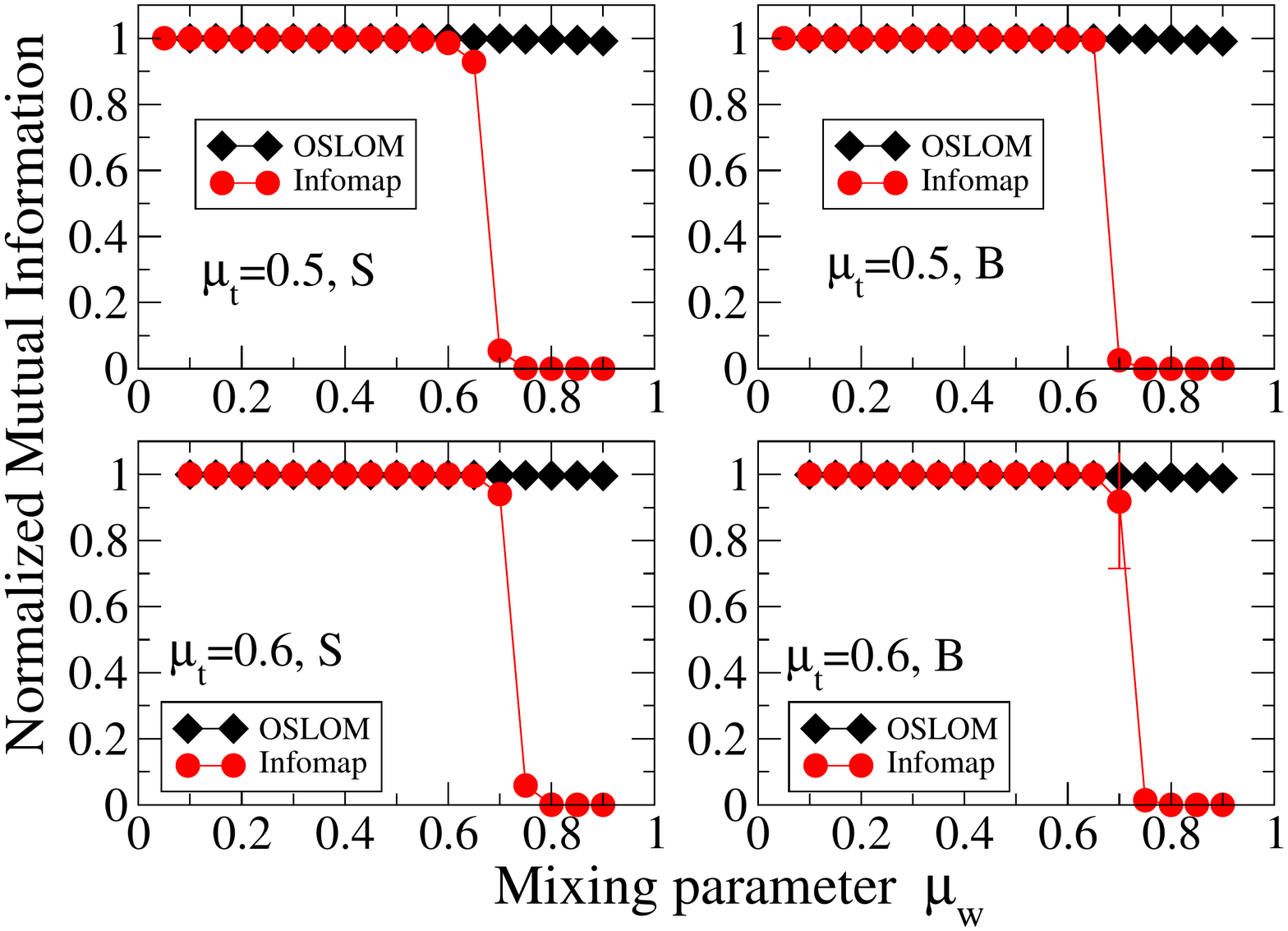}
\includegraphics[width=\columnwidth]{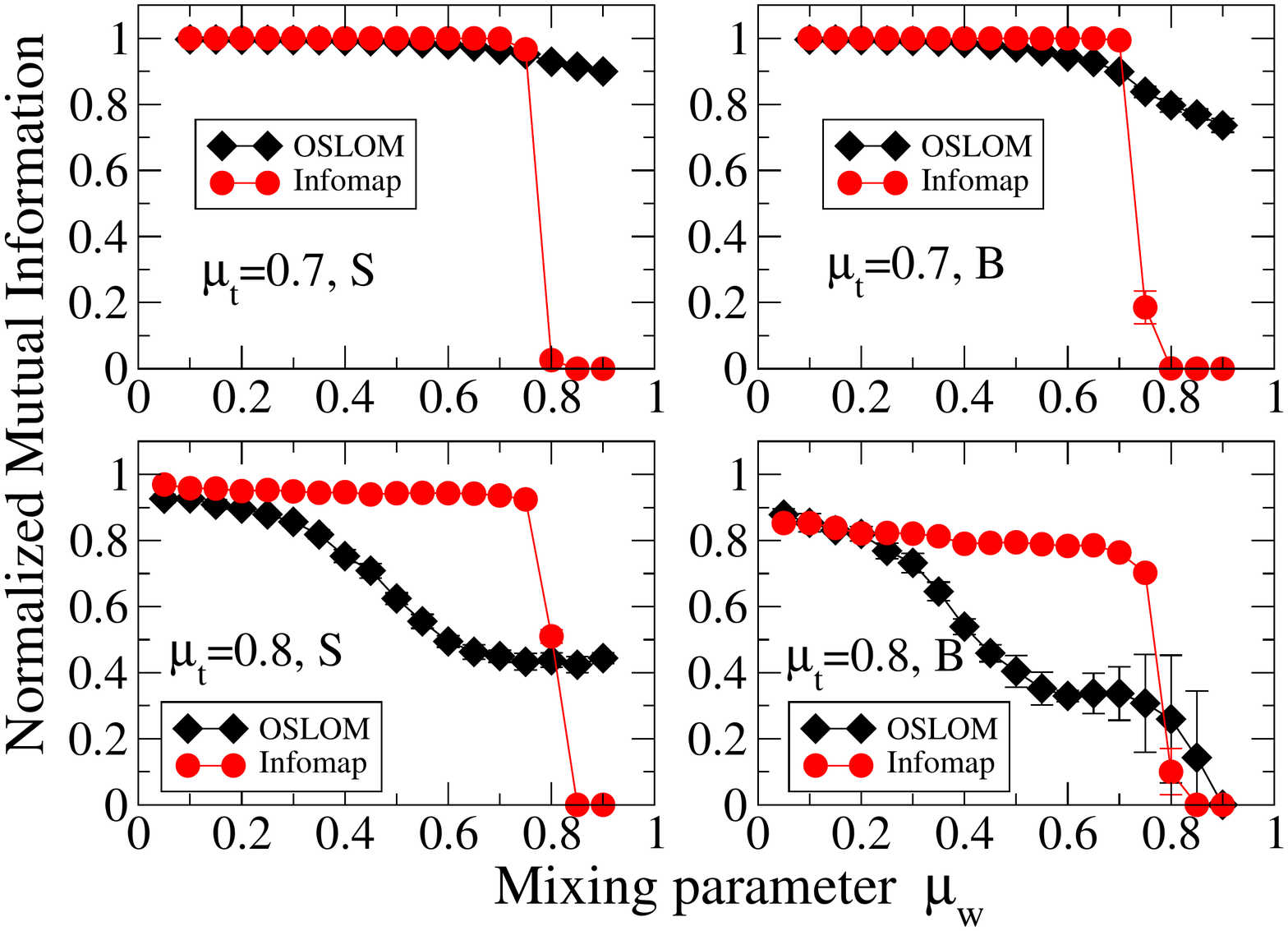}
\caption{Test on weighted LFR benchmark graphs (undirected and without
  overlapping communities).  The parameters are: $N=5000$, $\langle
k\rangle=20$, $k_{max}=50$, $\tau_1=2$,
$\tau_2=1$, $\beta=1.5$. Each panel corresponds to a given value of
the topological mixing parameter $\mu_t$ and of the community range (S
or B).}
\label{fig10}
\end{center}
\end{figure}
The benchmark by Girvan and Newman~\cite{girvan02} (GN benchmark) is a class of
graphs with $128$ vertices, each, divided into four equal-sized
groups. Every vertex has expected degree $16$ (with a very peaked
distribution about $16$). The (average) number of neighbors of a vertex within
its group is $k_{in}$, whereas the (average) number of external
neighbors is $k_{out}$. By construction, $k_{in}+k_{out}=16$. 
In the language of the planted $\ell$-partition model~\cite{condon01}, the
probability that a vertex is linked to another vertex of its group is
$p=k_{in}/31$, the probability that a vertex is linked to external
vertices is $q=k_{out}/96$. The condition $p>q$ for the four groups to
be communities is then equivalent to $k_{out} \lesssim 12$ (this does
not account for random fluctuations, though~\cite{bianconi09,lancichinetti10}).

Fig.~\ref{fig4} shows the Normalized Mutual Information (in the version devised in Ref.~\cite{lancichinetti09})
between the planted partition of the GN benchmark and the partition
found by the algorithm as a function of $k_{out}$. 
As a term of comparison we used again Infomap~\cite{rosvall08}.
Fig.~\ref{fig4} shows that Infomap is more accurate for low values of
$k_{out}$ than OSLOM, but its performance drops rapidly for
$k_{out}\gtrsim 6$, whereas OSLOM shows a slower decay.

OSLOM is slightly worse than Infomap because it finds several homeless
vertices, as we explained in the main text (Section~\ref{lfr}). 

\subsection{Weighted LFR benchmark}
\label{lfrw_app}

\begin{figure}
\begin{center}
\includegraphics[width=\columnwidth]{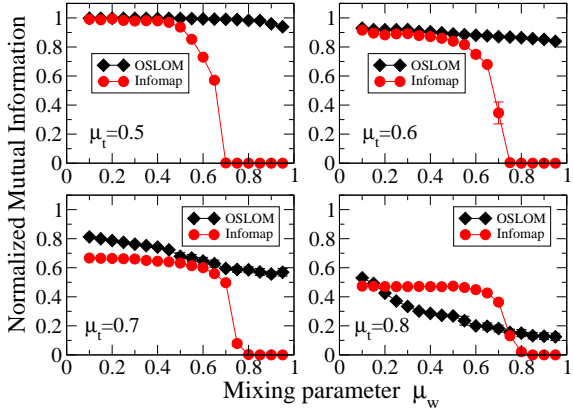}
\caption{Test on weighted LFR benchmark graphs (undirected and without
  overlapping communities). The parameters are: $N=50000$, $\langle
k\rangle=20$, $k_{max}=200$, $\tau_1=2$,
$\tau_2=1$, $\beta=1.5$. Each panel corresponds to a given value of
the topological mixing parameter $\mu_t$. The range of community sizes is
$[20, 1000]$.}
\label{fig11}
\end{center}
\end{figure}

In Figs.~\ref{fig10} and \ref{fig11} we report the comparative 
analysis of OSLOM and
Infomap on weighted LFR graphs. 
To build the weighted benchmark 
graphs~\cite{lancichinetti09b} one needs two
additional parameters: the exponent $\beta$ of the relation between
the strength of a vertex and its degree (the strength of a vertex is the sum of the weights
of the edges incident on the vertex); the weighted mixing parameter
$\mu_w$, which is the natural extension to weighted networks of the topological $\mu$ (that
here we call $\mu_t$), i.e. it is the ratio between the sum of
the weights on the edges joining a vertex to its neighbors in
different communities and the strength of the vertex. In the analysis, 
we fix the value of the
topological mixing parameter $\mu_t$ and see how the normalized mutual
information varies as a function of $\mu_w$. In Fig.~\ref{fig10}
the benchmark graphs consist of $5000$ vertices, and we consider the
usual two ranges of community sizes (S and B). In
Fig.~\ref{fig11} the graphs consist of $50000$ vertices, and we
consider a single, but much wider, range of community sizes (from $20$ to $1000$). 
When $\mu_t=0.5$ or $\mu_t=0.6$, we find that OSLOM detects the right clusters for
any value of $\mu_w$, for $N=5000$, which is truly
remarkable, while Infomap is unable to find the partition for
$\mu_w\gtrsim 0.6$. OSLOM's striking result comes from the fact that
the score $r_{tw}$ of a vertex on weighted graphs is given by the product of
two numbers, the topological score ${r'}_t$ and the weight score $r_w$
(Section~\ref{wnetworks}). If $\mu_t$ is not too large, the
topological term ${r'}_t$ is very low and brings down the whole score
$r_{tw}$, which remains significant for any choice of the weighted mixing
parameter $\mu_w$. Basically, OSLOM is able to recognize the right clusters 
from the topology alone.
When $\mu_t=0.5$ or $\mu_t=0.6$ and $N=50000$, OSLOM maintains an
excellent performance for the whole range of $\mu_w$, while Infomap again fails for
$\mu_w\gtrsim 0.6$. 
For $\mu_t=0.7$ the performances of the two algorithms worsen and
OSLOM is still superior, though the results are essentially comparable
for both network sizes. For
$\mu_t=0.8$ Infomap is more accurate than OSLOM, when $N=5000$, while
both methods are not very good when $N=50000$.
However, from Figs.~\ref{fig10} and \ref{fig11} it is apparent
that OSLOM works the better, the larger the network size. So, on very
large networks ($N\gg 50000$) we expect that OSLOM has a comparable or
superior performance than Infomap for every pair of values ($\mu_t$, $\mu_w$).
We also infer that the performance of both algorithms worsens if clusters are on average
larger.

\subsection{Directed LFR benchmark}
\label{lfrd_app}

\begin{figure}
\begin{center}
\includegraphics[width=\columnwidth]{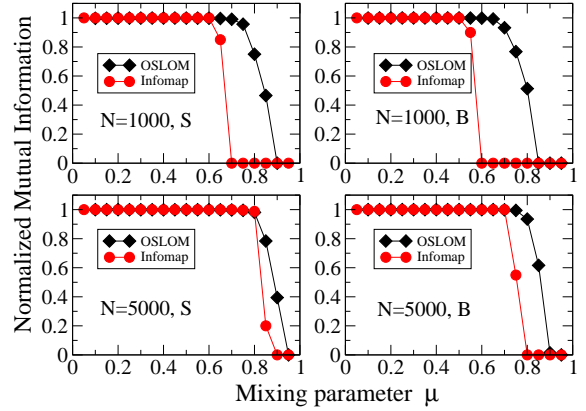}
\caption{Test on directed LFR benchmark graphs (unweighted and without
  overlapping communities). The parameters are: $\langle
k\rangle=20$, $k_{max}=50$, $\tau_{in}=2$,
$\tau_2=1$. Each panel corresponds to a given network size ($N=1000,
5000$) and community range (S or B). The mixing parameter $\mu$ refers to in-degree.}
\label{fig12}
\end{center}
\end{figure}

\begin{figure}
\begin{center}
\includegraphics[width=\columnwidth]{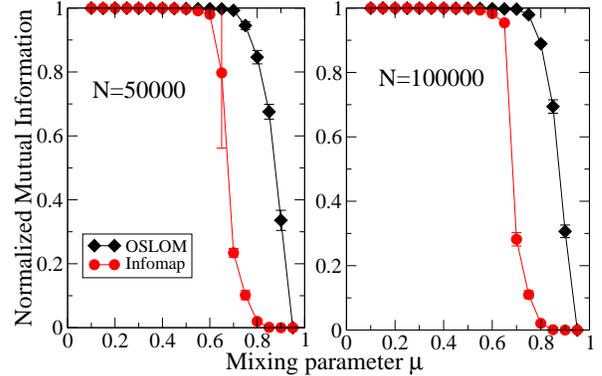}
\caption{Test on directed LFR benchmark graphs (unweighted and without
  overlapping communities). The parameters are: $\langle
k\rangle=20$, $k_{max}=200$, $\tau_{in}=2$,
$\tau_2=1$. We consider two large network sizes: $N=50000$ (left) and
$N=100000$ (right). The range of community sizes is
$[20, 1000]$. The mixing parameter $\mu$ refers to in-degree.}
\label{fig13}
\end{center}
\end{figure}

Figs.~\ref{fig12} and \ref{fig13} show the results of the test on directed LFR
graphs~\cite{lancichinetti09b}. 
This time we have to distinguish 
between in-degree (number of
incoming edges) and out-degree (number of outgoing edges) of a vertex.
The in-degree distribution is taken to be a power law, with exponent
$\tau_{in}$, whereas the out-degree is the same for all vertices, for
simplicity. The mixing parameter $\mu$ expresses the ratio of the
number of in-neighbors of a vertex belonging to different clusters and
the total number of in-neighbors of the vertex. The in-neighbor of a
vertex $i$ is any vertex $j$ connected to $i$ by an edge going from
$j$ to $i$. Figs.~\ref{fig12} and \ref{fig13} tell us that
OSLOM outperforms Infomap, especially when communities span a broader
range of sizes. The performances of both algorithms slightly worsen on
larger networks.

\section{Real-world systems}
\label{real_app}

\subsection{Zachary karate club}

The famous karate club network of
Zachary~\cite{zachary77} is a standard benchmark in community
detection. 
Vertices are members of a karate club in the United States, who were
monitored during a period of three years. Edges connect members
who had social interactions outside the club. After some time, a 
conflict between the club president and the instructor caused the fission of the club in two separate 
groups, supporting the instructor and the president, respectively.
\begin{figure}
\begin{center}
\includegraphics[width=1.1\columnwidth] {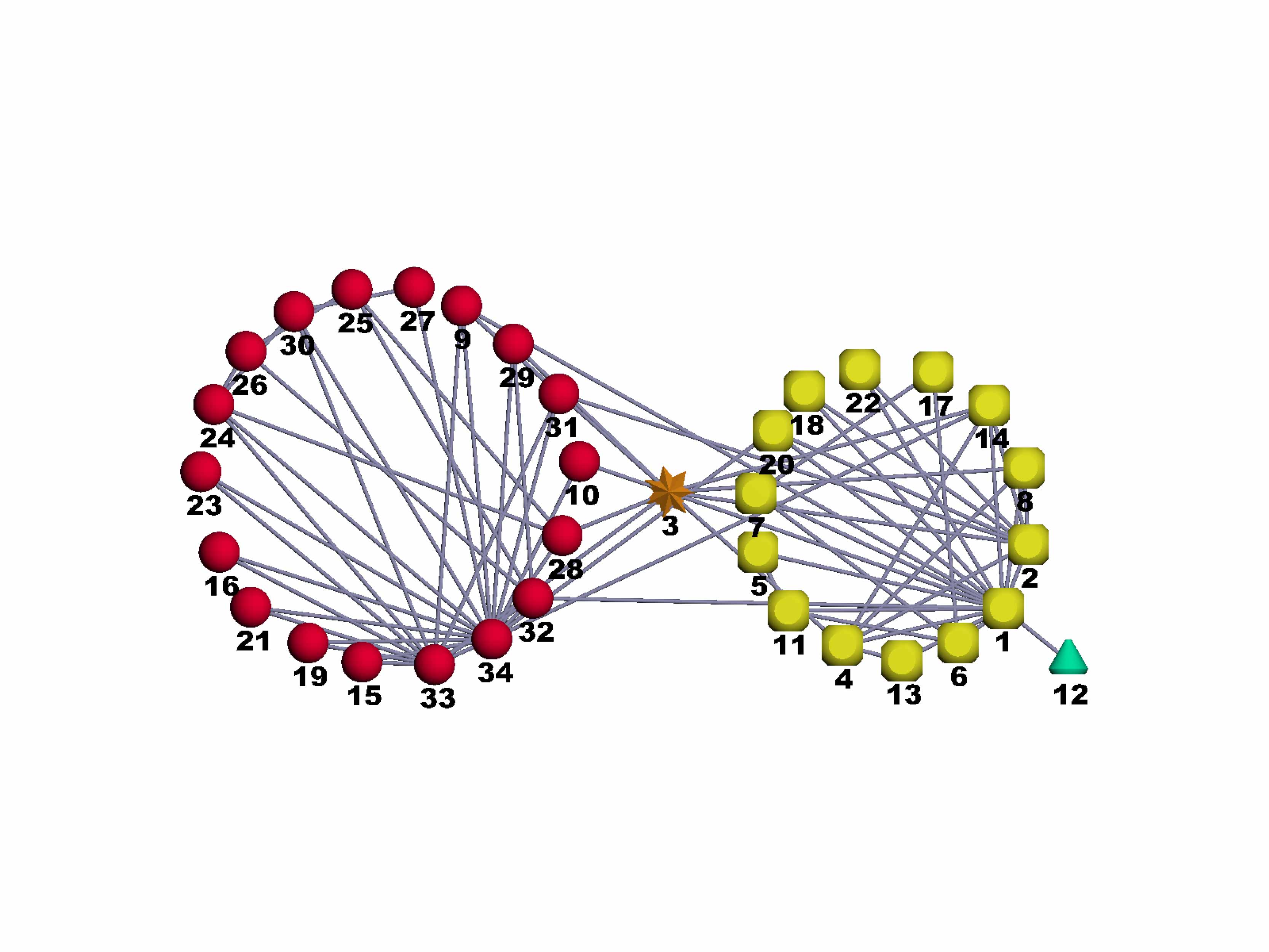}
\caption{Application of OSLOM to real networks: Zachary's karate club.}
\label{fig16}
\end{center}
\end{figure}
\begin{figure}
\begin{center}
\includegraphics[width=1.1\columnwidth]{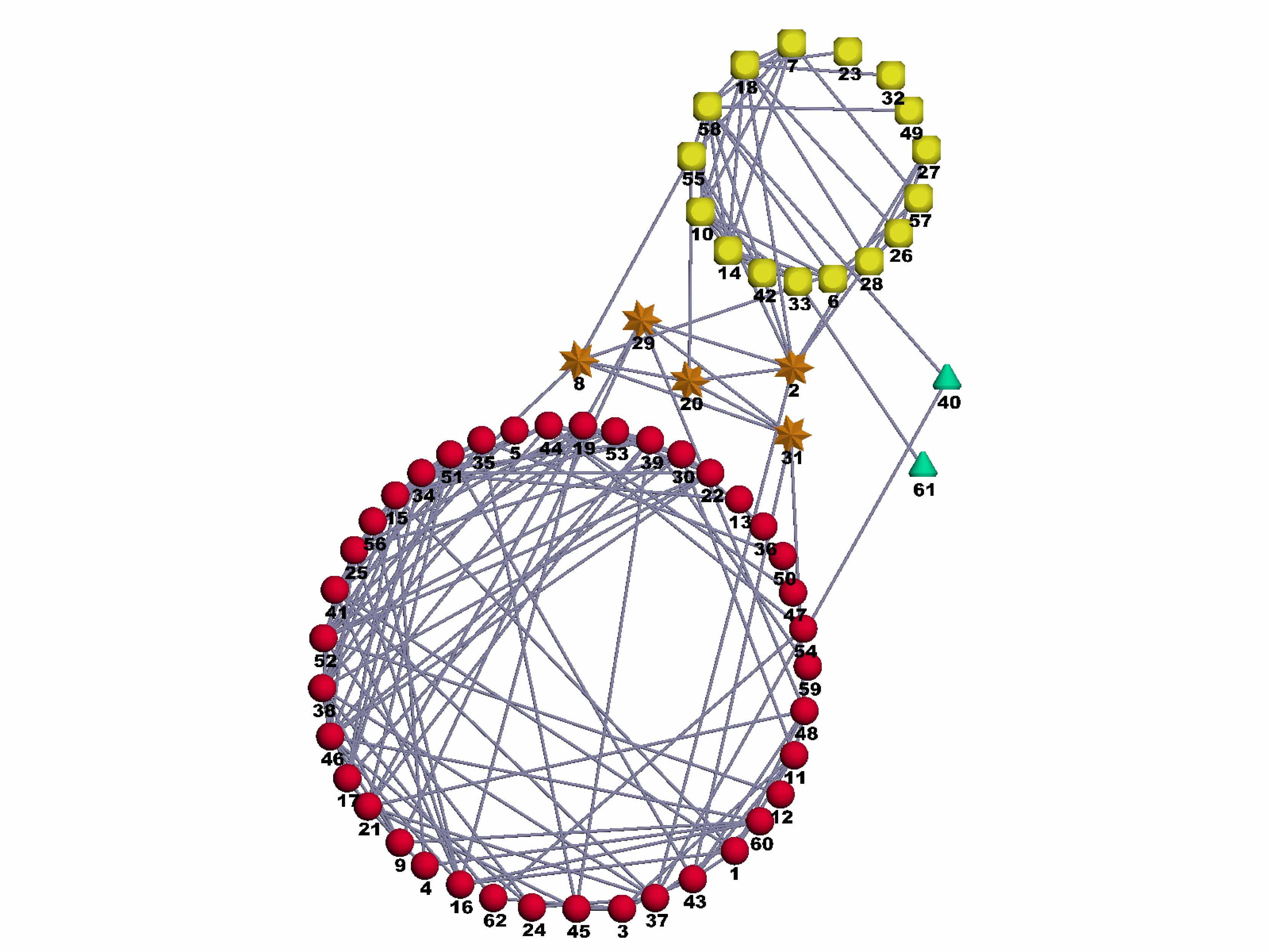}
\caption{Application of OSLOM to real networks: Lusseau's social
  network of bottlenose dolphins.}
\label{fig17}
\end{center}
\end{figure}
In Fig.~\ref{fig16} we see the community structure found by OSLOM. It
indeed finds two communities, plus a homeless vertex ($12$). Vertex
$3$ is shared between the two clusters, as it has several neighbors
in both groups. We shall illustrate
overlapping and homeless vertices with stars and triangles,
respectively. 
The communities coincide with the ones observed by Zachary with the
exception of vertices $3$ and $12$, which Zachary put with the
squares. However, vertex $3$ is overlapping, so it belongs to both
clusters, which seems quite reasonable by looking at the figure. Also,
vertex $12$ is homeless due to its loose relationship with its group
(it has only one neighbor). 

\subsection{Dolphin social network}

Fig.~\ref{fig17} presents OSLOM's results for 
the network of bottlenose dolphins living in Doubtful Sound
(New Zealand). The network was compiled by Lusseau~\cite{lusseau03}. 
Vertices of the network are dolphins and two dolphins are
connected if they were seen together more 
often than expected by chance. 
The dolphins separated in two groups
after one of them left the place for some time. OSLOM finds two
communities, with five overlapping vertices ($2$, $8$, $20$, $29$,
$31$), plus two homeless
vertices ($40$, $61$), which are  very loosely connected to the
rest of the graph. All vertices which are uniquely assigned to the same group
(indicated by the same symbol, square or circle, in the figure) are
classified in the same community by Lusseau as well.

\subsection{American college football}

Another well known benchmark in community detection is the network of 
American college football teams, compiled by 
Girvan and Newman~\cite{girvan02}. 
It comprises $115$ vertices, representing 
Division I-A colleges. Edges correspond
to games played by the teams against each other 
during the regular season of fall $2000$. 
The teams are divided into $12$ conferences. 
Games between teams in the same 
conference are usually (but not always)
 more frequent than games between teams of different conferences, 
so there is a organization in clusters where communities correspond to 
conferences.
\begin{figure*}
\begin{center}
\includegraphics[width=\textwidth]{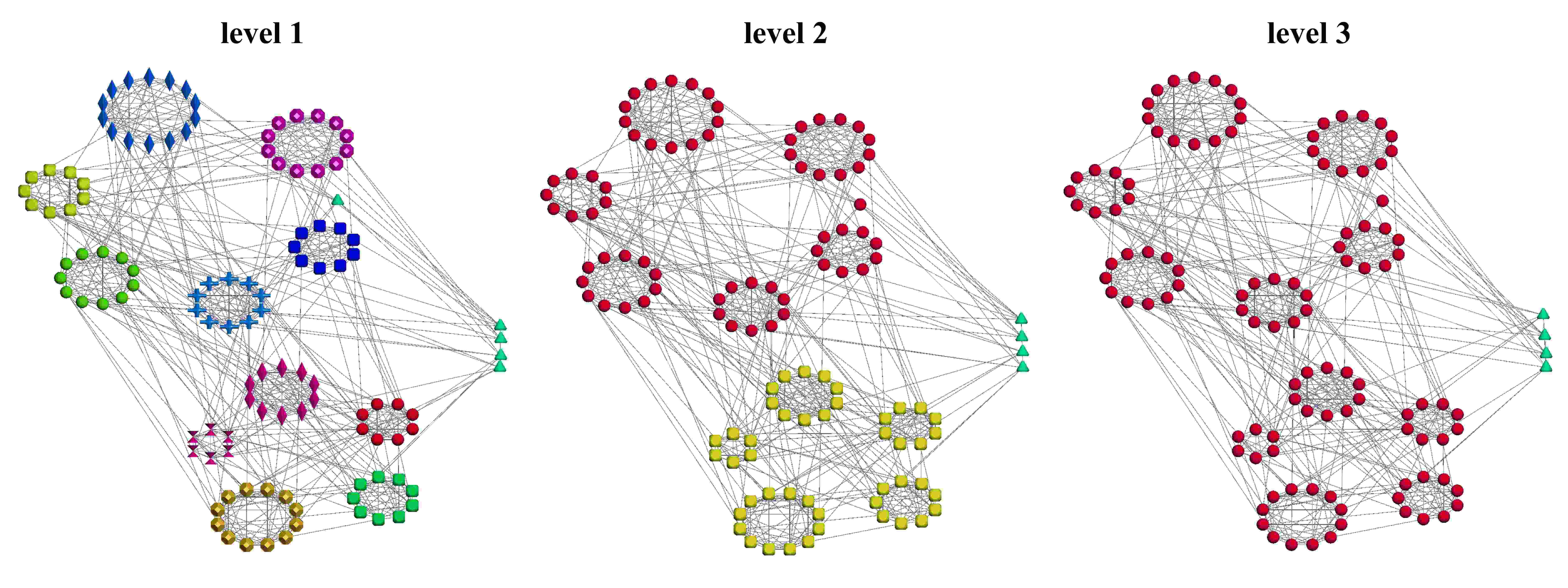}
\caption{Application of OSLOM to real networks: American college
  football network.}
\label{fig18}
\end{center}
\end{figure*}
\begin{figure*}
\begin{center}
\includegraphics[width=\textwidth]{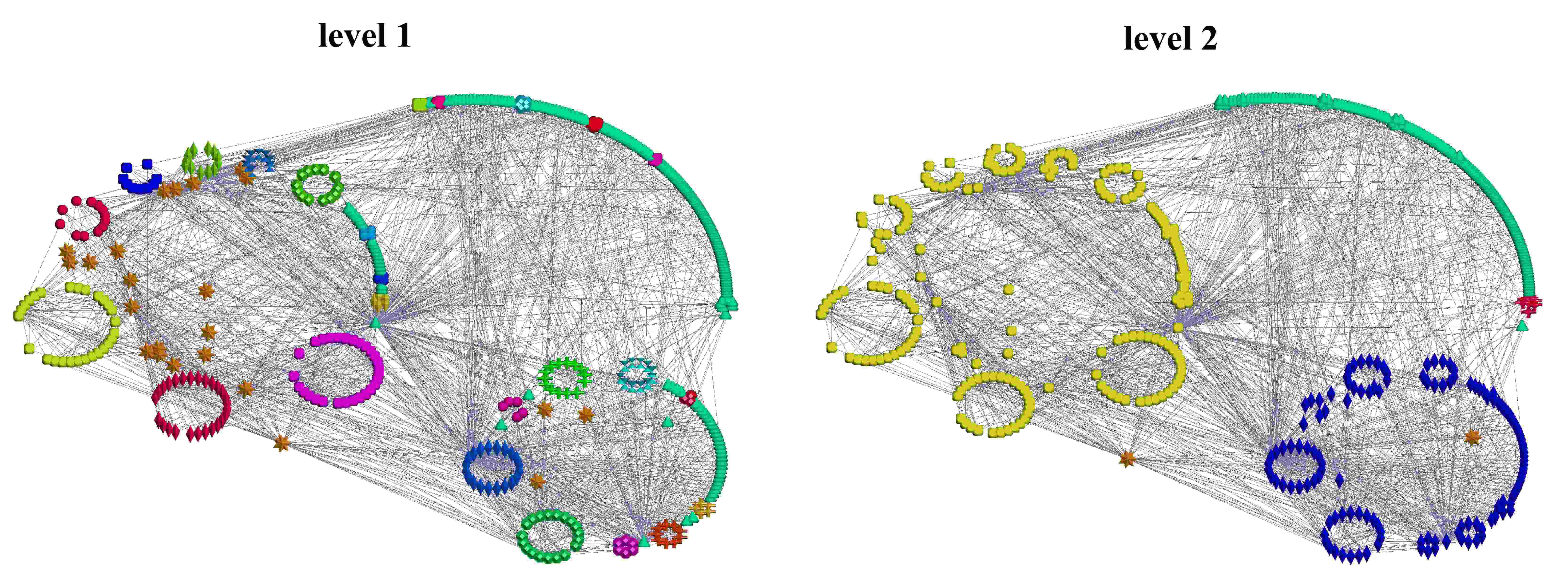}
\caption{Application of OSLOM to real networks: metabolic network of
  {\it C. elegans}.}
\label{fig19}
\end{center}
\end{figure*}
In Fig.~\ref{fig18} we see that OSLOM finds three
hierarchical levels. The lowest level consists of $11$ clusters and
$5$ homeless vertices. There are no overlapping vertices. 
Six clusters correspond exactly
to the conferences, three others match the conferences up to 
one vertex, one up to two vertices, 
the last cluster along with the homeless vertices mostly mix teams of the
conferences Sun Belt and Independents. The latter is not a proper
conference, whereas Sun Belt includes colleges which are
geographically very spreadout, so they happen to play quite often
games with the other teams, resulting much more mixed with them than
teams of other conferences. Interestingly, in the second hierarchical
level we find two large communities (plus four homeless teams), 
corresponding quite well to a geographical separation of the 
colleges in East and West. 

\subsection{{\it C.~elegans} metabolic network}

Fig.~\ref{fig19} presents the community structure of the metabolic
network of {\it C. elegans}. 
The network has been compiled by Duch and
Arenas~\cite{duch05} and it has been often used in applications of
community detection algorithms. Vertices are metabolites and edges
connect pairs of metabolites involved in at least one biochemical
reaction. OSLOM finds two hierarchical levels, the lower with $25$
clusters, the higher with $3$ (but one of them is much smaller than the
other two). The fraction of
homeless vertices in the lower level is larger than $20\%$ (see Table~1) and the network appears rather
``noisy''.
\clearpage

\end{appendix}

%

\end{document}